\documentclass[a4paper,fleqn]{cas-dc}

\usepackage[utf8]{inputenc}
\usepackage[T1]{fontenc}

\usepackage{amsmath}
\usepackage{amssymb}
\usepackage{bm}

\usepackage{enumitem}
\usepackage{url}
\usepackage{tipa}
\usepackage{newunicodechar}
\usepackage{textcomp}

\usepackage{booktabs}
\usepackage{multirow}
\usepackage{array}
\usepackage{tabularx}
\usepackage{makecell}
\usepackage{stfloats}

\usepackage{tikz}
\usepackage{pgfplots}
\usepgfplotslibrary{groupplots}
\pgfplotsset{compat=1.18}

\usepackage[numbers,sort&compress]{natbib}

\usepackage{algorithm}
\usepackage{algpseudocode}

\newcommand{\tablefontsize}{\footnotesize}

\usepackage{hyperref}
\hypersetup{hidelinks}

\begin{document}

\let\WriteBookmarks\relax
\def\floatpagepagefraction{1}
\def\textpagefraction{.001}


\shorttitle{Adaptive and High-Fidelity Matching Algorithm for Retrieval-Based Accent Conversion}

\shortauthors{Wang et al.}

\title[mode=title]
{Adaptive and High-Fidelity Matching Algorithm for Retrieval-Based Accent Conversion}


\tnotemark[1]

\tnotetext[1]{
This work was supported by the National Natural Science Foundation of China
[grant numbers 61771196 and 61872143].
}


\author[1]{Xinran Wang}

\author[1]{Ning Chen}
\cormark[1]

\author[1]{Yixin Luo}

\author[1]{Hongqing Zhu}

\author[1]{Li Tian}


\affiliation[1]{
    organization={School of Information Science and Engineering,
    East China University of Science and Technology},
    city={Shanghai},
    postcode={200237},
    country={China}
}


\cortext[cor1]{
Corresponding author. E-mail address:
\href{mailto:nchen@ecust.edu.cn}{nchen@ecust.edu.cn}
}


\begin{abstract}
Accent conversion (AC) aims to transform accent-related acoustic characteristics of the source speech to a target accent while preserving linguistic content and speaker identity. Unlike conventional generative AC models, the retrieval-based ones, which employ $k$-nearest neighbors ($k$NN) to directly retrieve acoustic features from target-accent speech and so do not need source-accent speech during training, are more effective. However, existing retrieval-based AC frameworks are limited by three fundamental issues: the limited intelligibility of the converted speech resulted from the incomplete phoneme coverage achieved by sequential selection-based matching pool construction, the limited adaptability to the variations in the source speech content, and low spectral continuity resulted from coarse-grained style modulation. To address these limitations, we propose Anisotropy-Aware Adaptive $k$-Nearest Neighbor Accent Conversion (AA-$k$NN-AC), an enhanced retrieval-based AC model. First, we design a phoneme coverage-based matching pool construction strategy, employing greedy selection with diversity-aware scoring to achieve balanced and near-complete phonetic representation under limited target accent data. Second, we propose an anisotropy-aware adaptive $k$NN retrieval mechanism that dynamically determines the neighborhood size according to the singular value decomposition (SVD)-based anisotropy analysis, enhancing its adaptability to the variations in the source speech. Third, we present a progressive Adaptive Instance Normalization (AdaIN)-based fine-grained decoder by integrating temporal up-sampling within AdaIN blocks, enhancing spectral continuity of the converted speech. Experimental results on the L2-ARCTIC and VCTK datasets demonstrate that the proposed method outperforms existing retrieval-based and generative
baselines in both objective and subjective evaluations, achieving high intelligibility, speaker similarity, and naturalness. Demo page is available at \url{https://aa-knn-ac.github.io/demo/}.

\end{abstract}



\maketitle
\section{Introduction}

Accent conversion (AC) aims to transform accent-related characteristics in speech, such as pronunciation patterns and prosody, to match a target accent, while preserving the linguistic content and the source speaker's voice identity~\cite{zhao2019accent, ding2022accentron}. The potential applications of this technology range from computer-assisted language learning to cross-cultural communication and media localization~\cite{jin2023voice, zhong2025accentbox}. Despite substantial progress, available systems still struggle to simultaneously achieve high conversion fidelity, robust generalization to unseen speakers, and preservation of fine-grained acoustic details.

\begin{figure}
    \centering
    \includegraphics[width=\linewidth]{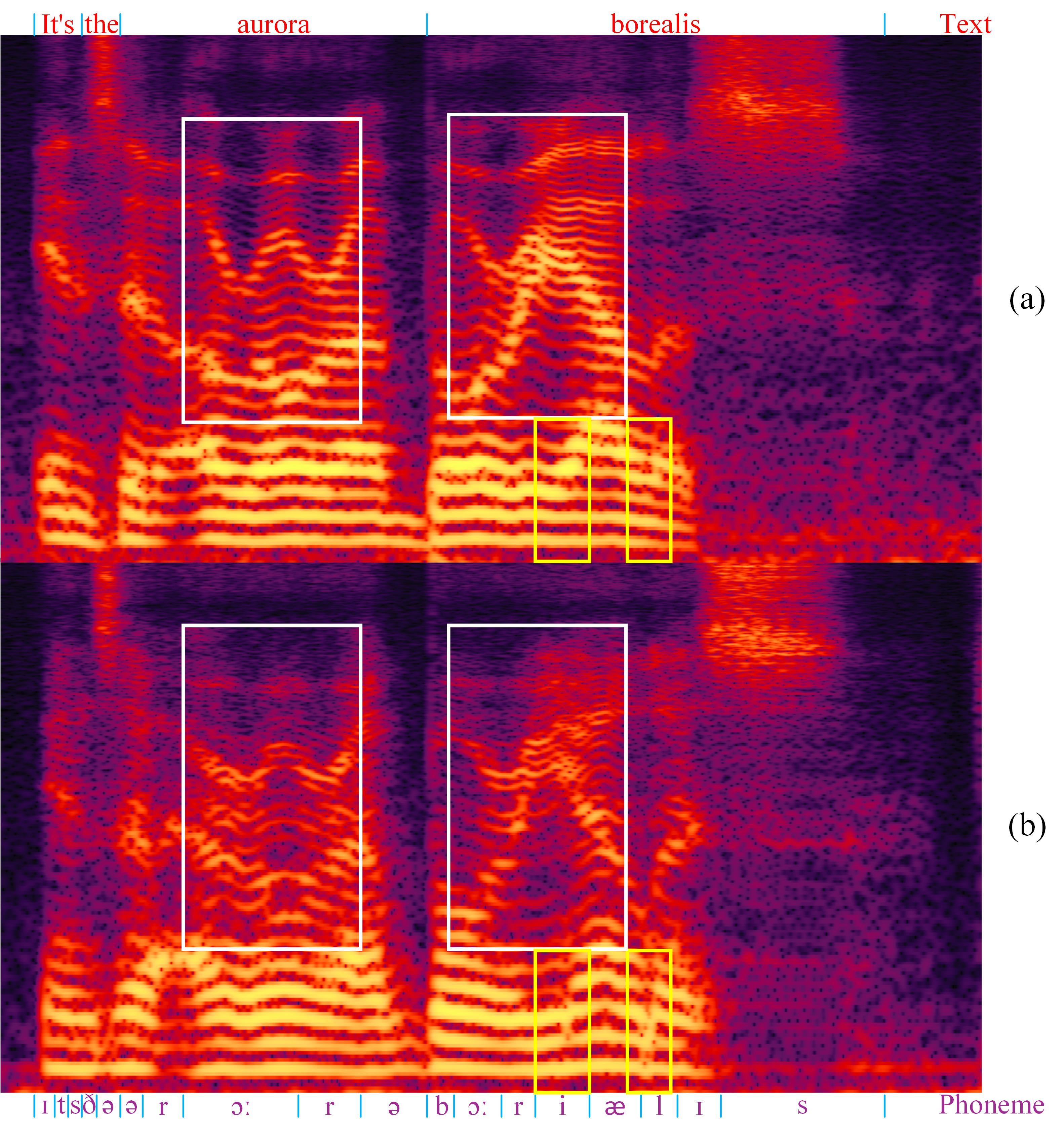}
    \caption{Mel-spectrograms of speech generated by (a) the proposed method and (b) $k$NN-AC for the utterance ``It's the aurora borealis.'' White boxes highlight vowel regions where $k$NN-AC exhibits over-smoothed high-frequency harmonic details due to the mismatch between coarse-grained style injection and fine-grained temporal reconstruction. Yellow boxes indicate spectral discontinuities in $k$NN-AC caused by inaccurate retrieval resulting from insufficient phoneme coverage and the use of fixed $k$ values in $k$NN search.}
    \label{fig:knnac_limitation}
\end{figure}

The evolution of AC systems is characterized by a progressive relaxation of data constraints. Early approaches relied on parallel training paradigms,  requiring paired speech corpora from the same speaker across multiple accents~\cite{aryal2015reduction}, but the prohibitive cost of data collection limited their practical applicability. To overcome this, research shifted toward non-parallel approaches using intermediate representations, e.g., phonetic posteriorgrams (PPGs)~\cite{zhao2019accent, bai2025accent}, autoencoder-based methods~\cite{saito2021nonparallel, qian2019autovc, chan2022speechsplit, wu2020vqvc}, and benchmark challenges~\cite{lorenzotrueba2018vcc, yi2020vcc}. However, these methods typically operate under a closed-set assumption, where the target speaker must be observed during training. Recent self-supervised learning (SSL)-based AC methods further relax data constraints by leveraging SSL models, such as HuBERT and WavLM, to extract robust content representations, enabling approaches such as feature disentanglement~\cite{jin2023voice, liu2020end, ma2024accentvits} and Multi-Level VAEs~\cite{melechovsky2024accent, zhou2023optimization}. Nevertheless, a fundamental challenge persists: empirical studies have shown that parametric waveform generation may produce over-smoothed outputs with attenuated high-frequency details, as models tend to favor average spectral patterns when mapping content representations to speech signals~\cite{wang2021accent, zhao2021converting}.

Retrieval-based methods offer a paradigm shift to circumvent this bottleneck. Rather than generating acoustic features, these approaches directly retrieve them from real speech data, thereby sidestepping the generalization limitations of parametric modeling. Pioneered by $k$-nearest neighbors ($k$NN)-VC~\cite{baas2023voice}, this paradigm operates on the premise that for any phonetic unit in the source speech, a perceptually similar unit exists in the target speaker's data pool. By retrieving and concatenating these nearest-neighbor feature frames, the synthesized speech preserves the natural texture and spectral details of actual recordings. Inspired by this work, Luo et al.~\cite{yixin2025zero} proposed the $k$NN-AC model, a retrieval-based framework that leverages WavLM representation to identify acoustically similar content segments from target accent data, followed by a dual-stage Adaptive Instance Normalization (AdaIN) Vocoder to progressively align retrieved features with the target speaker's acoustic characteristics.

Despite its effectiveness, the $k$NN-AC framework exhibits several structural limitations, as illustrated in Fig.~\ref{fig:knnac_limitation}. First, the sequential selection-based matching pool construction lacks systematic coverage, which leads to insufficient rare phonemes and low phoneme coverage. Second, the use of a fixed $k$ in $k$NN retrieval fails to adapt to local feature geometry. In regions where feature representations vary rapidly across neighboring frames, large $k$ values may introduce mismatched neighbors, whereas in regions with more uniform feature distributions, small $k$ values can lead to unstable retrieval. These two limitations jointly manifest as spectral discontinuities, as evidenced by the yellow boxes in Fig.~\ref{fig:knnac_limitation}(b), where abrupt spectral transitions occur at phoneme boundaries (e.g., the /r/-/i/ and /æ/-/l/ transitions in ``borealis'') due to retrieval of mismatched acoustic units, whereas our method in Fig.~\ref{fig:knnac_limitation}(a) demonstrates smoother and more natural spectral continuity. Third, style modulation is performed at a fixed resolution. The dual-stage AdaIN decoder injects style statistics at coarse temporal scales and relies on subsequent HiFi-GAN layers for resolution recovery, leading to a mismatch between style modulation and fine-grained temporal reconstruction. As shown in the white boxes of Fig.~\ref{fig:knnac_limitation}(b), the vowel segments in ``aurora'' and ``borealis'' exhibit attenuated high-frequency harmonic structures compared to our method in Fig.~\ref{fig:knnac_limitation}(a), indicating the loss of fine-grained spectral details during the decoupled up-sampling process.

To address these limitations, we propose an enhanced retrieval-based AC framework with the following three improvements: First, for matching pool construction, we design a phoneme coverage-based strategy, which employs greedy selection with diversity-aware scoring to ensure balanced and near-complete coverage of phonetic units under the Zipfian distribution of speech. Second, for adaptive retrieval, we propose an anisotropy-aware adaptive $k$NN (AA-$k$NN) strategy, which leverages singular value decomposition (SVD)-derived condition numbers~\cite{klema1980singular} to dynamically adjust the neighborhood size according to local feature-space geometry. Such design mitigates mismatched retrieval in anisotropic regions while maintaining stability in smoother regions. Third, for vocoder refinement, we design a progressive up-sampling mechanism, which integrates transposed convolutions within AdaIN blocks, explicitly coupling style injection with resolution recovery in a coarse-to-fine manner. Such alignment enables more consistent temporal reconstruction and reduces spectral discontinuities.

The main contributions of the proposed model, named AA-$k$NN-AC, can be summarized as follows:
\begin{enumerate}[leftmargin=*, label=(\roman*)]
\item A phoneme coverage-based matching pool construction strategy, which addresses the incomplete phoneme coverage issue through greedy optimization with diversity-aware scoring, is designed to enhance the intelligibility of the converted speech;

\item An anisotropy-aware adaptive $k$NN retrieval strategy, which improves the adaptability to the variations in the source speech content by dynamically adjusting neighborhood size based on the anisotropy of local features, is proposed;

\item A progressive AdaIN-based fine-grained decoder, which performs style modulation on speech contents at different sampling rates, is designed to enhance spectral continuity of the converted speech;

\item Extensive experiments are conducted on both VCTK and L2-ARCTIC datasets to demonstrate the competitive performance of the proposed model in both objective and subjective evaluations, and the effectiveness of each of the above designs.
\end{enumerate}
\section{Related Work}

\subsection{Generative Accent Conversion}

AC aims to modify accent-related pronunciation and prosodic characteristics while preserving linguistic content and speaker identity. Early AC systems were mainly developed under parallel or carefully aligned settings, where source- and target-accent utterances shared the same linguistic content. Representative examples include statistical parametric and articulatory-synthesis-based approaches that explicitly modeled accent-related acoustic or articulatory transformations~\cite{aryal2015reduction}. Although such methods benefit from direct supervision, their reliance on parallel or speaker-dependent data limits scalability.

To reduce the reliance on parallel training data, later studies explored phonetic posteriorgram (PPG)-based conversion~\cite{zhao2019accent,zhao2019using,zhao2021converting}, which introduced phonetic representations as intermediate features for accent modification. More recent generative AC systems further improved flexibility through disentanglement, adversarial learning, variational modeling, token-based modeling, and synthesis-guided training, including Accentron~\cite{ding2022accentron}, VP-AC~\cite{jin2023voice}, disentanglement-based methods~\cite{wang2021accent,huang2024usd,melechovsky2024accent}, discrete-token approaches~\cite{bai2025accent}, and text-to-speech (TTS)/VITS-based approaches~\cite{zhou2023tts,zhou2024accented,chen2024transfer,ma2024accentvits,nguyen2025improving,zhong2025accentbox}. Recent continuous-time generative paradigms, such as diffusion-based speech conversion systems represented by Seed-VC~\cite{liu2024}, further improve modeling flexibility and generation quality.

Despite these advances, generative AC systems generally rely on learned model parameters to reconstruct target-accent acoustic characteristics. This generation-dominant paradigm provides strong modeling flexibility, but it may make the preservation of fine-grained target-accent details challenging, especially for high-frequency harmonic structures, transient phonetic regions, and subtle spectral variations. When such details are not accurately reconstructed, the converted speech may exhibit over-smoothed acoustic patterns or reduced spectral fidelity.

\subsection{Retrieval-Based Accent Conversion}

Retrieval-based speech conversion provides an alternative to fully parametric generation by reusing acoustic representations extracted from real speech. A representative example is $k$NN-VC~\cite{baas2023voice}, which performs voice conversion by retrieving nearest-neighbor target-speaker features from a reference pool constructed in a self-supervised representation space. Such representations, including wav2vec 2.0 and WavLM~\cite{baevski2020wav2vec,chen2022wavlm}, have been widely used in speech representation learning and downstream speech conversion tasks.

Compared with generative AC systems, retrieval-based methods introduce target-accent information by directly selecting acoustic representations from real target-accent speech, rather than reconstructing all target-accent details solely from model parameters. Since the retrieved features originate from real recordings, this paradigm helps preserve natural acoustic texture and fine-grained spectral structures. In this sense, retrieval-based AC reduces the burden of parametric reconstruction and provides a more direct way to incorporate target-accent acoustic characteristics.

This retrieval-based idea was later extended to accent conversion in $k$NN-AC~\cite{yixin2025zero}, which retrieves target-accent content features and combines them with source-speaker information through an AdaIN-based decoder. Compared with discrete-token, disentanglement-based, TTS/VITS-based, and diffusion-based AC systems~\cite{bai2025accent,wang2021accent,huang2024usd,melechovsky2024accent,jin2023voice,zhou2023tts,zhou2024accented,chen2024transfer,ma2024accentvits,nguyen2025improving,liu2024}, retrieval-based AC introduces target-accent information directly from real target-accent speech before waveform reconstruction, rather than relying solely on parametric generation.

Despite its promise, existing retrieval-based AC methods still face several limitations. First, target-accent matching pools are often constructed through random, chronological, or fixed-duration selection, which does not guarantee balanced phoneme coverage and may lead to mismatched retrieval for rare phonemes. Second, existing $k$NN-based systems typically use a fixed neighborhood size, although speech representations vary substantially across stable and rapidly changing regions. As a result, fixed-$k$ retrieval may either oversmooth transient regions or become unstable in homogeneous regions. Third, coarse-resolution style modulation may limit spectral continuity, because the downstream vocoder must recover fine-grained harmonic structures from relatively coarse conditioning representations.

To address the above issues, we adopt $k$NN-AC~\cite{yixin2025zero} as the baseline framework and introduce three corresponding improvements: phoneme-coverage-aware matching-pool construction, anisotropy-aware adaptive $k$NN retrieval, and progressive AdaIN-based fine-grained decoding.

\section{Method}

\begin{figure*}[t]
    \centering
    \includegraphics[width=0.85\textwidth]{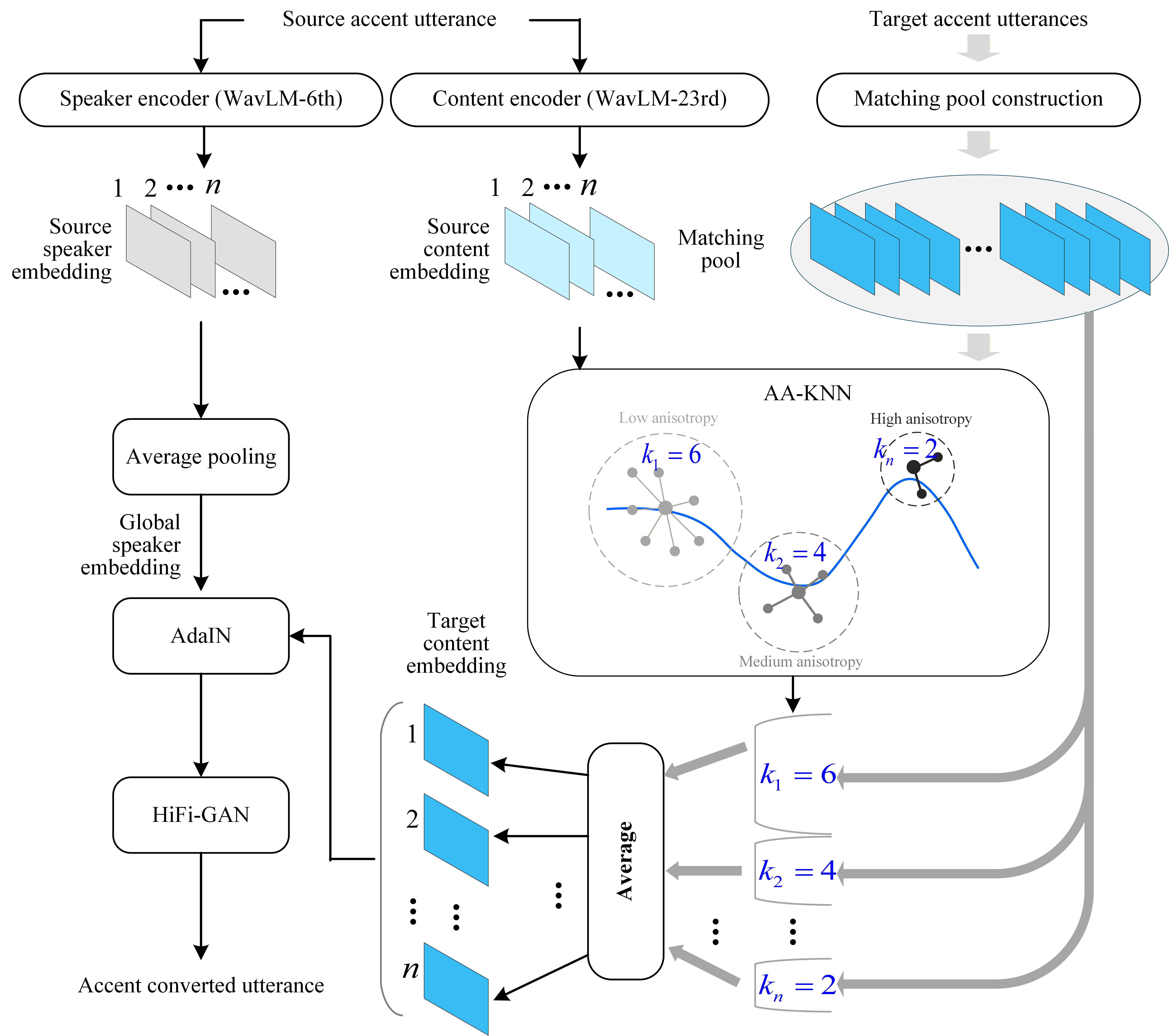}
    
    \caption{The Inference pipeline of the proposed model. The source utterance is disentangled into speaker and content embeddings using the 6th and 23rd layers of WavLM, respectively. The AA-$k$NN module retrieves target content features from the matching pool by dynamically adjusting the neighbor count $k$ within the range of $[2, 6]$ based on the local anisotropy of the feature space. Finally, the synthesized target content embedding and the source speaker embedding are fused via the AdaIN module to generate the converted speech using HiFi-GAN.}
    \label{fig:framework}
\end{figure*}

Inspired by the $k$NN-based voice conversion model and AC model~\cite{baas2023voice}~\cite{yixin2025zero}, we propose a $k$NN-based AC model (see Fig.~\ref{fig:framework}). Given that the entanglement between accent and linguistic content is stronger than that between timbre and content, the performance of the $k$NN-based AC model relies more heavily on the construction of the matching pool and the matching strategy than the $k$NN-based VC model. To this end, a carefully designed matching-pool construction strategy, together with an adaptive matching strategy based on Anisotropy-Aware Adaptive $k$NN (AA-$k$NN), is proposed. In addition, guided by layer-wise probing studies~\cite{chen2022wavlm, pasad2021layerwise}, the 6th and 23rd layers of WavLM-large\footnote{\url{https://huggingface.co/patrickvonplaten/wavlm-libri-clean-100h-large}} are adopted as the speaker and content embeddings, respectively. Specifically, Layer 6 is used as the speaker embedding because middle WavLM layers retain relatively strong speaker- and timbre-related information~\cite{chen2022wavlm}. Layer 23 is used as the content embedding because the penultimate layer provides a content-dominant representation while avoiding the stronger task-specific bias of the final layer~\cite{pasad2021layerwise}. Since accent cues can be partially entangled with linguistic content in SSL representations, we do not claim that this layer pair is universally optimal; rather, it is adopted as a principled design choice supported by prior layer-wise probing analyses. A HiFi-GAN vocoder is adopted to generate the converted speech.

\subsection{Matching Pool}

In previously proposed $k$NN-based VC and AC models~\cite{baas2023voice,yixin2025zero}, sequential selection-based matching-pool construction is typically adopted. However, because phonemes are unevenly distributed in natural speech, sequential selection may overrepresent common phonemes while leaving rare phonemes underrepresented or absent in the constructed matching pool. This imbalance can degrade retrieval quality and compromise accent conversion performance. To address this issue, we develop a CMU dictionary-based text-to-phoneme mapping mechanism and define a comprehensive quality evaluation function for quantitative sample assessment (see Eq. (1)):
\begin{equation}
Q(x_k) = \lambda_1\mathcal{U}_k + \lambda_2H(x_k) + \lambda_3\mathcal{D}_k + \lambda_4R_k + \lambda_5E_k
\end{equation}
where $\mathcal{U}_k$ denotes the number of unique phonemes, $\mathcal{D}_k$ counts the broad phoneme categories, $R_k$ is the rare phoneme count, $H(x_k)$ is the shannon diversity, and $E_k = H(x_k)/\log_2(\mathcal{U}_k)$ represents the distribution evenness. Here, $\lambda_1$--$\lambda_5$ are the corresponding coefficients. $H(x_k)$ is defined in Eq. (2).
\begin{equation}
H(x_k) = -\sum_p \pi_p(x_k) \log_2 \pi_p(x_k)
\end{equation}
where $\pi_p(x_k)$ denotes the frequency of phoneme $p$ in sample $x_k$. The coefficients in Eq. (1) are empirically assigned to emphasize rare phoneme coverage and phoneme-level diversity, while retaining phoneme uniqueness, broad-category coverage, and distribution evenness as complementary constraints. The robustness of this weighting strategy is further examined in the ablation study.

This method adopts a three-stage selection strategy. In the first stage, speech samples containing rare phonemes, such as /\textipa{Z}/, /\textipa{T}/, /\textipa{D}/, and /\textipa{N}/, are prioritized. In each search iteration, the sample that contains these rare phonemes and achieves the highest comprehensive quality score $Q(x_k)$ is selected, ensuring mandatory coverage of low-frequency phonemes. In the second stage, samples are selected in descending order of $Q(x_k)$ to supplement phoneme coverage and improve distribution uniformity. Finally, in the third stage, samples are chosen based on the quality efficiency per unit time, $Q(x_k)/t_k$, until the predefined total duration is reached.

After completing speech sample selection, a pre-trained WavLM model is employed to extract features from the selected samples. The output from the 23rd layer is adopted as the content embedding for matching. The anisotropy scores are computed independently for each selected utterance, and the resulting utterance-level features and anisotropy scores are then concatenated to construct the matching-pool retrieval index.

\subsection{AA-\texorpdfstring{$k$}{k}NN}

In traditional $k$NN-based accent conversion methods, all source content embeddings are matched using a fixed number of neighbors. However, such a fixed-$k$ strategy fails to adapt to the varying geometric complexity across different regions of the feature space. In locally structured, directionally strong regions (high anisotropy), a larger $k$ value introduces irrelevant neighbors and reduces matching accuracy. Conversely, in smooth and homogeneous regions (low anisotropy), a smaller $k$ value leads to unstable matching. To address this issue, we propose an adaptive $k$NN matching model based on local feature anisotropy. We employ SVD to quantify the local variation of the feature distribution. As pointed out by Klema and Laub~\cite{klema1980singular}, directly computing the eigenvalues of the covariance matrix unnecessarily introduces squared terms, amplifying numerical errors, whereas SVD operates directly on the original matrix and circumvents this problem. Furthermore, Levinson et al.~\cite{levinson2020analysis} demonstrated that SVD orthogonalization is optimal in the least-squares sense, and that singular values exhibit favorable stability against data perturbations~\cite{klema1980singular}. As illustrated in Fig.~\ref{fig:framework}, we calculate an anisotropy score based on SVD to dynamically adjust the neighborhood size $k$ for each source content embedding.

\subsubsection{Anisotropy Score Estimation}

First, the input utterance is segmented into $L$ frames and the feature of the $j$-th frame is $\bm{f}_j \in \mathbb{R}^{d \times 1}$. Assuming that the receptive field size is $w$, the feature matrix corresponding to $\bm{f}_j$, denoted as $\tilde{\bm{F}}_j \in \mathbb{R}^{d \times (w_l + w_r + 1)}$, can be obtained with Eq. (3):
\begin{equation}
\tilde{\bm{F}}_j = [\bm{f}_{j-w_{jl}}, \dots, \bm{f}_{j+w_{jr}}]
\end{equation}
where $w_{jl} = \min(j-1, \lfloor w/2 \rfloor)$ and $w_{jr} = \min(L - j, \lfloor w/2 \rfloor)$ are left range and right range of the receptive field of $\bm{f}_j$.

Second, as for SVD, $\tilde{\bm{F}}_j$ is centered with Eq. (4) to obtain $\hat{\bm{F}}_j$.
\begin{equation}
\hat{\bm{F}}_j = \tilde{\bm{F}}_j \left( \bm{I}_{w_j} - \frac{1}{w_j} \bm{1}_{w_j} \bm{1}_{w_j}^T \right)
\end{equation}
where $w_j = w_l + w_r + 1$, and $\bm{I}_{w_j}$ is the identity matrix of size $w_j$. Then SVD is performed on $\hat{\bm{F}}_j$ via Eq. (5):
\begin{equation}
\hat{\bm{F}}_j = \bm{U}_j \bm{\Sigma}_j \bm{V}_j^T
\end{equation}
where $\bm{\Sigma}_j=\mathrm{diag}(a_1,a_2,\dots,a_Y)$ contains the singular values in descending order, and $\bm{U}_j$ and $\bm{V}_j$ denote the left and right singular-vector matrices, respectively. Further, to ensure numerical stability, a self-adaptive threshold $\tau_j$ is calculated with Eq. (6) and employed to filter out small singular values:
\begin{equation}
\tau_j = \max \left( a_1 \cdot \epsilon_{\text{base}}, a_1 \cdot \frac{\epsilon_{\text{ada}}}{\sqrt{w_j}} \right)
\end{equation}
where $\epsilon_{\text{base}} = 10^{-6}$ and $\epsilon_{\text{ada}} = 10^{-2}$ are the base relative threshold and the adaptive factor, respectively.

Third, the anisotropy score, denoted as $\alpha_j$, is calculated with Eq. (7):
\begin{equation}
\alpha_j = \ln \left( \frac{\max(S_j)}{\min(S_j)} \right)
\end{equation}
where $S_j = \{ a_m \mid a_m > \tau_j \}$ is the set of significant singular values.

While, for the frames located at the edges of the utterance, since $w_j < 2 \lfloor w/2 \rfloor + 1$, the self-adaptive threshold may become larger and thus some small informative singular values will be suppressed. As a result, an underestimated anisotropy will be obtained.

To solve this issue, the anisotropy score is corrected with Eq. (8) to obtain $\hat{\alpha}_j$.
\begin{equation}
\hat{\alpha}_j = \alpha_j \cdot \sqrt{(2\lfloor w/2 \rfloor + 1)/w_j}
\end{equation}
where the factor $\sqrt{(2\lfloor w/2 \rfloor + 1)/w_j}$ is introduced as an empirical boundary correction, which is motivated by the threshold shown in Eq. (6).

\subsubsection{Adaptive Neighborhood Size Calculation and Matching}

Assuming that $\bm{f}_i^q, i=1,\dots,L^q$ is the feature of the $i$-th frame of the query utterance, its anisotropy score, denoted as $\hat{\alpha}_i^q$, can be calculated with above anisotropy score calculation method. Then, its adaptive neighborhood size can be determined as follows.

First, the percentile rank, denoted as $\rho_i^q$, is calculated with Eq. (9):
\begin{equation}
\rho_i^q = \frac{1}{L} \sum_{j=1}^L \mathbb{I} \left[ \hat{\alpha}_j \le \hat{\alpha}_i^q \right]
\end{equation}
where $\mathbb{I}[\cdot]$ is an indicator function, which returns 1 if the condition is true and 0 otherwise.

Second, considering that the frame with higher anisotropy score requires smaller neighborhood size, while lower anisotropy score requires larger neighborhood size, the adaptive neighborhood size, denoted as $k_i^{\text{ada}}$, is calculated with Eq. (10):
\begin{equation}
k_i^{\text{ada}} = k_{\min} + \lfloor (1 - \rho_i^q) \cdot (k_{\max} - k_{\min}) + 0.5 \rfloor
\end{equation}
where $k_{\max}$ and $k_{\min}$ are manually set maximum and minimum values of neighborhood sizes, respectively.

Furthermore, to make $k_i^{\text{ada}}$ suitable for the frames located at the boundaries of the utterance, it is further modified with Eq. (11) to obtain $\tilde{k}_i^{\text{ada}}$:
\begin{equation}
\tilde{k}_i^{\text{ada}} = \begin{cases}
\max(k_{\min}, k_i^{\text{ada}} - \Delta k_{\text{bnd}}), 
& i \le \lfloor w/2 \rfloor \text{ or} \\
& i \ge L^q - \lfloor w/2 \rfloor + 1 \\
k_i^{\text{ada}}, & \text{otherwise}
\end{cases}
\end{equation}
where $\Delta k_{\text{bnd}}$ is boundary reduction value.

Third, the cosine distance between $\bm{f}_i^q$ and each of $\bm{f}_j, j=1,\dots,L$, denoted as $\text{dis}(i,j)$, is calculated with Eq. (12):
\begin{equation}
\text{dis}(i,j) = 1 - \frac{(\bm{f}_i^q)^T \bm{f}_j}{\Vert\bm{f}_i^q\Vert_2 \Vert\bm{f}_j\Vert_2}
\end{equation}

Finally, the synthesized feature of the $i$-th frame in the query utterance, denoted as $\hat{\bm{f}}_i^q$, is obtained with Eq. (13):
\begin{equation}
\hat{\bm{f}}_i^q = \frac{1}{\tilde{k}_i^{\text{ada}}} \sum_{m=1}^{\tilde{k}_i^{\text{ada}}} \bm{f}_m^*
\end{equation}
where $\bm{f}_m^*$ is the $m$-th nearest neighbor of $\bm{f}_i^q$ in the matching pool, denoted as $\bm{F} = \{\bm{f}_1, \dots, \bm{f}_L\}$.

Then the content embedding of the query utterance, denoted as $\bm{C}^q$, can be obtained via
\begin{equation}
\bm{C}^q = [\hat{\bm{f}}_1^q, \hat{\bm{f}}_2^q, \dots, \hat{\bm{f}}_{L_q}^q],
\end{equation}
which serves as the input of the subsequent modules.

\subsection{Progressive AdaIN-based Fine-grained Decoder}

\begin{figure}
    \centering
    \includegraphics[width=\linewidth]{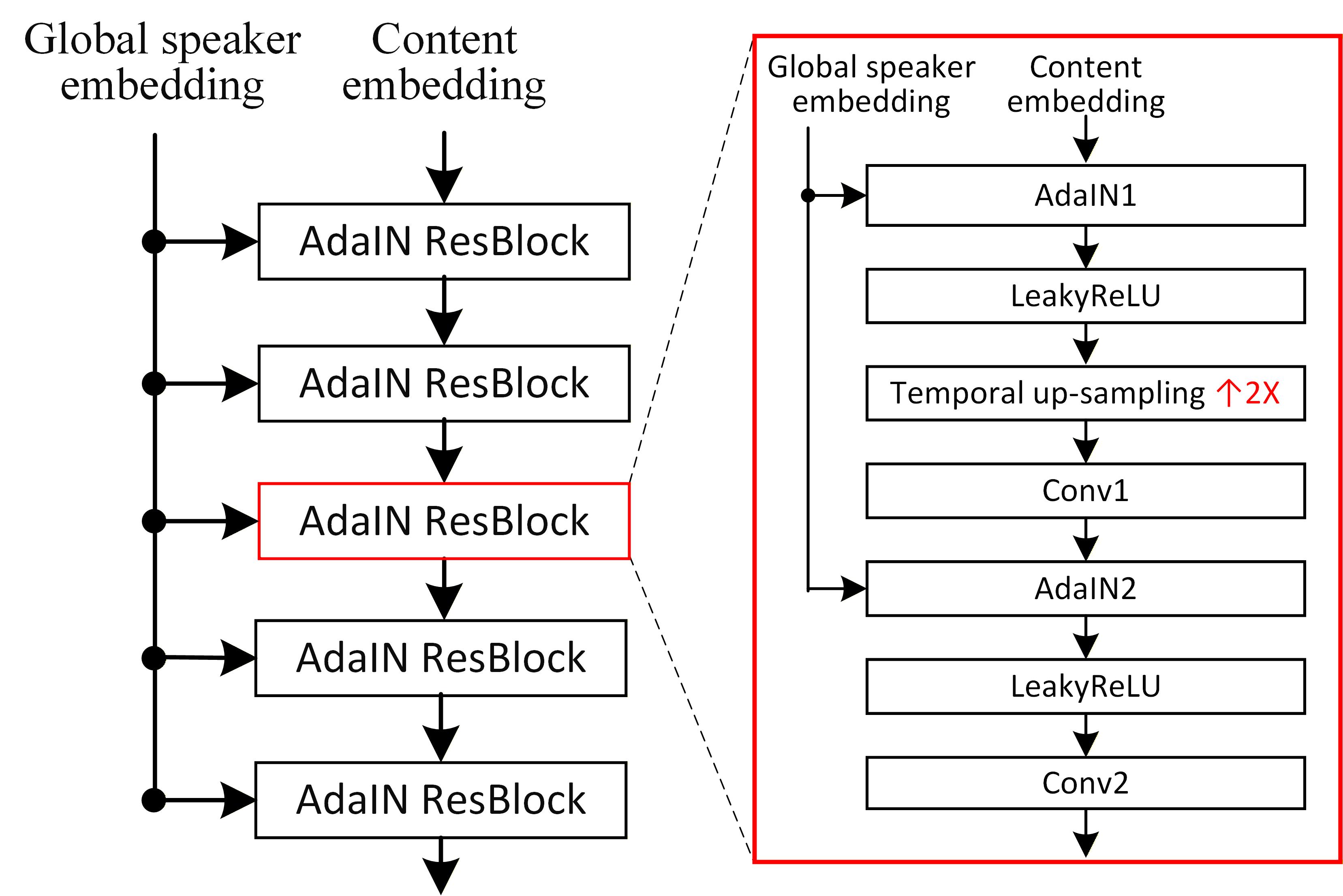} 
    \caption{Architecture of the proposed speaker-content integration module. \textbf{Left:} The cascade of AdaIN ResBlocks fusing the speaker embedding $\bm{s}$ and content embedding $\bm{C}$. \textbf{Right:} Detailed structure of the up-sampling AdaIN ResBlock, where a 2$\times$ temporal up-sampling layer is placed between two AdaIN modulation layers.}
    \label{fig:adain_resblock}
\end{figure}

The conventional $k$NN-AC framework~\cite{yixin2025zero} performs speaker-content modulation at a single coarse temporal resolution before waveform up-sampling, which may limit fine-grained spectral reconstruction. To address this issue, we propose a progressive AdaIN-based decoder that couples speaker-content modulation with temporal up-sampling. As shown in Fig.~\ref{fig:adain_resblock}, the decoder processes a global speaker embedding $\bm{s}$ from the 6th layer of WavLM and a content embedding $\bm{C}$ from AA-$k$NN through five cascaded AdaIN ResBlocks.

For 16 kHz audio with a 20 ms frame shift, the WavLM feature sequence has a temporal resolution of 50 Hz, corresponding to 320 waveform samples per frame. In the conventional $k$NN-AC vocoder~\cite{yixin2025zero}, the 50 Hz content embeddings are directly converted to waveform using upsample\_rates: [10, 8, 2, 2]. In our architecture, one 2$\times$ up-sampling operation is moved from the HiFi-GAN vocoder into the AdaIN module, and the remaining vocoder up-sampling rates are changed to [10, 4, 2, 2]. Thus, the overall up-sampling factor remains 320, while speaker-content modulation is performed at both 50 Hz and 100 Hz temporal resolutions.

The third AdaIN ResBlock implements this dual-scale modulation by placing a 2$\times$ temporal up-sampling layer between two AdaIN layers. The first AdaIN layer modulates the representation at 50 Hz, and the second one refines the up-sampled representation at 100 Hz.

The AdaIN operation aligns the channel-wise statistics of $\bm{C}$ with affine parameters derived from $\bm{s}$:
\begin{equation}
\text{AdaIN}(\bm{C}, \bm{s}) = (1 + \bm{\gamma}(\bm{s})) \odot \frac{\bm{C} - \mu(\bm{C})}{\text{std}(\bm{C})} + \bm{\beta}(\bm{s})
\label{eq:adain_principle}
\end{equation}
where $\odot$ denotes element-wise multiplication, while $\mu(\bm{C})$ and $\mathrm{std}(\bm{C})$ denote the per-channel mean and standard deviation computed over the temporal dimension. The scale $\bm{\gamma}(\bm{s})$ and bias $\bm{\beta}(\bm{s})$, generated via linear projections of $\bm{s}$, modulate the normalized target-accent content features according to the source speaker embedding.

\section{Experimental Setup}

\subsection{Datasets}
The datasets adopted in the experiment are as follows (see Table~\ref{tab:datasets}):

\subsubsection{Training Data}
The decoder and HiFi-GAN vocoder are trained on the train-clean-100 subset of LibriSpeech~\cite{panayotov2015librispeech}, which contains approximately 100 hours of read speech from 251 native American English speakers. 

\subsubsection{Target Accent Data}
The target-accent matching pool for the proposed model is constructed using 25~minutes of recordings from speaker BDL (male) in the CMU-ARCTIC database~\cite{kominek2004cmu}. This pool is constructed offline before inference and kept fixed for all conversion experiments. These recordings are used only to construct the American-accent matching pool for retrieval and are not used to update model parameters. During conversion, the system only takes the source utterance as input; no additional target speech is needed, and the matching pool will not be rebuilt or updated for different source speakers.

\subsubsection{Source Accent Data}
To evaluate the generalization of the proposed model across diverse foreign accents, three non-native speakers are selected from two established corpora: ASI (Hindi accent) and ABA (Arabic accent) from L2-ARCTIC~\cite{zhao2018l2}, and P248 (Indian accent) from VCTK\footnote{\url{https://datashare.ed.ac.uk/handle/10283/3443}}. All recordings are resampled to 16~kHz to maintain uniformity.

\begin{table*}[!t]
\centering
\caption{Details of datasets}
\label{tab:datasets}
\footnotesize
\setlength{\tabcolsep}{6pt}
\renewcommand{\arraystretch}{1.08}
\begin{tabular}{@{}lllll@{}}
\toprule
\textbf{Datasets} & \textbf{Role} & \textbf{Speaker (Gender)} & \textbf{Accent / L1} & \textbf{Duration (h)} \\
\midrule
LibriSpeech~\cite{panayotov2015librispeech} & Training & 251 speakers & American & 100.0 \\
CMU-ARCTIC~\cite{kominek2004cmu} & Target accent & BDL (M) & American & 1.0 \\
CMU-ARCTIC~\cite{kominek2004cmu} & Target accent & SLT (F) & American & 1.0 \\
\midrule
L2-ARCTIC~\cite{zhao2018l2} & Source accent & ASI (M) & Hindi (L1) & 1.0 \\
L2-ARCTIC~\cite{zhao2018l2} & Source accent & ABA (M) & Arabic (L1) & 1.1 \\
VCTK & Source accent & P248 (F) & Indian (Regional)\textsuperscript{*} & 0.4 \\
\bottomrule
\end{tabular}

\vspace{1mm}
\parbox{0.86\textwidth}{\footnotesize \textit{Note:} \textsuperscript{*}In the Speaker (Gender) column, M stands for male and F stands for female. For L2-ARCTIC, the accent is specified by the speaker's first language (L1). For VCTK, ``Indian'' denotes a broad regional English variety as per the official dataset metadata, with the specific L1 unspecified.}
\end{table*}

\subsection{Implementation Details}

A fixed 25-minute speech segment is used to construct the target-accent matching pool offline before inference. This pool is applied for all test utterances, and only the source speech is required during conversion. The anisotropy scores of the matching pool are pre-computed offline, which reduces the inference-time overhead of adaptive $k$NN retrieval. The frame length and frame shift are 25 ms and 20 ms, respectively. The key parameters of the proposed model are set as follows: the receptive field $w$ is set to 10, the neighborhood size $k$ is dynamically determined within the range $[2, 6]$ based on local anisotropy, and the boundary reduction amount $\Delta k_{\text{bnd}}$ is set to 1. The training is conducted on a single NVIDIA RTX 4090 GPU using the AdamW optimizer with a batch size of 64 and a learning rate of $2 \times 10^{-4}$. In all experiments, the coefficients in Eq.~(1) are empirically set to $\lambda_1=0.20$, $\lambda_2=0.30$, $\lambda_3=0.05$, $\lambda_4=0.40$, and $\lambda_5=0.05$.

\subsection{Baseline Systems}

The performance of the proposed model is compared with those of previously proposed AC models based on different strategies, including retrieval-based ones~\cite{baas2023voice,yixin2025zero}, discrete-unit generation-based one~\cite{bai2025accent}, native TTS priors-based ones~\cite{chen2024transfer,nguyen2025improving}, disentanglement-based ones~\cite{melechovsky2024accent,jin2023voice}, and a recent diffusion transformer-based framework (Seed-VC)~\cite{liu2024}. As detailed in Table~\ref{tab:baselines}, for the models that provide codes (denoted as `Reproducible'), such as~\cite{baas2023voice,yixin2025zero,bai2025accent,liu2024}, the performances are compared on L2-ARCTIC (ASI, ABA) and VCTK (P248), while for those that only provide demos (denoted as `Demos'), such as~\cite{jin2023voice,melechovsky2024accent,chen2024transfer,nguyen2025improving}, the performances are compared on the demo samples provided by the corresponding models.

\begin{table*}[!t]
\centering
\caption{Training and testing configurations of various models}
\label{tab:baselines}
\footnotesize
\setlength{\tabcolsep}{6pt}
\renewcommand{\arraystretch}{1.08}
\begin{tabular}{@{}llll@{}}
\toprule
\textbf{Categories} & \textbf{Models} & \textbf{Training data} & \textbf{Testing data} \\
\midrule

\multirow{4}{*}{Reproducible}
& $k$NN-VC~\cite{baas2023voice} & LibriSpeech (100h) & \multirow{4}{*}{\begin{tabular}[c]{@{}l@{}}L2-ARCTIC (ASI, ABA),\\ VCTK (P248)\end{tabular}} \\
& $k$NN-AC~\cite{yixin2025zero} & LibriSpeech (100h) & \\
& TokAN~\cite{bai2025accent} & LibriTTS (360h) + L2-ARCTIC & \\
& Seed-VC~\cite{liu2024} & Emilia-101k5 (101000h) & \\
\midrule

\multirow{4}{*}{Demos}
& MLVAE-ADV~\cite{melechovsky2024accent} & L2-ARCTIC + CMU-ARCTIC & \multirow{4}{*}{Various official demo samples} \\
& TTS-AC~\cite{chen2024transfer} & LibriTTS + LJSpeech & \\
& VITS-AC~\cite{nguyen2025improving} & L2-ARCTIC + Synthetic Data & \\
& VP-AC~\cite{jin2023voice} & VCTK + L2-ARCTIC & \\
\midrule

\multirow{2}{*}{\textbf{Proposed}}
& \multirow{2}{*}{\textbf{AA-$k$NN (Ours)}} 
& \multirow{2}{*}{LibriSpeech (100h)}
& L2-ARCTIC (ASI, ABA), VCTK (P248), \\
& & & Various official demo samples \\

\bottomrule
\end{tabular}
\end{table*}

\begin{table*}[!t]
\centering
\caption{Performance comparison with 'reproducible' baselines}
\label{tab:baseline1}
\tablefontsize
\setlength{\tabcolsep}{4pt}
\renewcommand{\arraystretch}{1.02}
\begin{tabular*}{\textwidth}{@{\extracolsep{\fill}}llcccccccc@{}}
\toprule
& & \multicolumn{5}{c}{\textbf{Objective Metrics}} 
& \multicolumn{3}{c}{\textbf{Subjective Metrics}} \\
\cmidrule(lr){3-7} \cmidrule(lr){8-10}
\textbf{Dataset} & \textbf{Model}
& \textbf{WER(\%)}$\downarrow$
& \textbf{CER(\%)}$\downarrow$
& \textbf{SECS}$\uparrow$
& \textbf{$\Delta$PPG}$\downarrow$
& \textbf{RTF}$\downarrow$
& \textbf{NMOS}$\uparrow$
& \textbf{SMOS}$\uparrow$
& \textbf{ACT}$\uparrow$ \\
\midrule

\multirow{5}{*}{L2 (ASI)~\cite{zhao2018l2}}
& $k$NN-VC~\cite{baas2023voice} 
& 38.84 & 26.57 & -- & 0.93 & \textbf{0.05} 
& $3.31\pm0.10$ & -- & $3.14\pm0.14$ \\

& $k$NN-AC~\cite{yixin2025zero} 
& 13.12 & 6.59 & 0.64 & 0.40 & 0.08 
& $3.81\pm0.09$ & $3.24\pm0.09$ & $3.92\pm0.09$ \\

& TokAN~\cite{bai2025accent} 
& 13.24 & 7.53 & \textbf{0.82} & 0.38 & 0.61 
& $\mathbf{4.24\pm0.09}$ & $3.62\pm0.12$ & $3.95\pm0.13$ \\

& Seed-VC~\cite{liu2024} 
& 9.58 & 4.72 & -- & 0.69 & 0.20 
& $4.07\pm0.14$ & -- & $3.70\pm0.23$ \\

& Ours 
& \textbf{8.35} & \textbf{4.04} & 0.80 & \textbf{0.36} & 0.13 
& $4.20\pm0.09$ & $\mathbf{3.74\pm0.09}$ & $\mathbf{4.18\pm0.08}$ \\

\midrule

\multirow{5}{*}{L2 (ABA)~\cite{zhao2018l2}}
& $k$NN-VC~\cite{baas2023voice} 
& 37.33 & 24.16 & -- & 0.91 & \textbf{0.07} 
& $2.97\pm0.14$ & -- & $3.24\pm0.13$ \\

& $k$NN-AC~\cite{yixin2025zero} 
& 17.52 & 8.47 & 0.69 & 0.53 & 0.11 
& $3.69\pm0.13$ & $3.16\pm0.13$ & $3.82\pm0.12$ \\

& TokAN~\cite{bai2025accent} 
& 18.94 & 11.40 & \textbf{0.82} & \textbf{0.46} & 0.63 
& $\mathbf{4.11\pm0.09}$ & $3.53\pm0.13$ & $\mathbf{4.09\pm0.12}$ \\

& Seed-VC~\cite{liu2024} 
& 16.04 & 7.54 & -- & 0.78 & 0.19 
& $4.04\pm0.11$ & -- & $3.85\pm0.13$ \\

& Ours 
& \textbf{15.72} & \textbf{7.31} & \textbf{0.82} & 0.51 & 0.15 
& $4.08\pm0.10$ & $\mathbf{3.82\pm0.14}$ & $3.98\pm0.10$ \\

\midrule

\multirow{5}{*}{VCTK (P248)}
& $k$NN-VC~\cite{baas2023voice} 
& 10.10 & 5.31 & -- & 0.72 & \textbf{0.08} 
& $3.51\pm0.13$ & -- & $3.66\pm0.12$ \\

& $k$NN-AC~\cite{yixin2025zero} 
& 6.40 & 2.71 & 0.75 & 0.32 & 0.10 
& $3.94\pm0.11$ & $3.34\pm0.11$ & $3.82\pm0.10$ \\

& TokAN~\cite{bai2025accent} 
& 24.74 & 15.62 & \textbf{0.88} & 0.46 & 0.50 
& $\mathbf{4.19\pm0.08}$ & $3.45\pm0.16$ & $3.97\pm0.14$ \\

& Seed-VC~\cite{liu2024} 
& 6.38 & 2.90 & -- & 0.55 & 0.18 
& $4.12\pm0.12$ & -- & $4.00\pm0.11$ \\

& Ours 
& \textbf{5.86} & \textbf{2.37} & 0.78 & \textbf{0.30} & 0.11 
& $4.16\pm0.09$ & $\mathbf{3.62\pm0.10}$ & $\mathbf{4.05\pm0.08}$ \\

\bottomrule
\end{tabular*}
\vspace{1mm}
\parbox{\textwidth}{\tablefontsize \textit{Note:} Bold indicates the best performance in each group; $\downarrow$ and $\uparrow$ indicate that lower and higher values are better, respectively.}
\end{table*}

\begin{table*}[!t]
\centering
\caption{Performance comparison with different `Demos' baselines}
\label{tab:baseline2}
\tablefontsize
\setlength{\tabcolsep}{4pt}
\renewcommand{\arraystretch}{1.03}
\begin{tabular*}{\textwidth}{@{\extracolsep{\fill}}llccccccc@{}}
\toprule
& & \multicolumn{4}{c}{\textbf{Objective Metrics}} 
& \multicolumn{3}{c}{\textbf{Subjective Metrics}} \\
\cmidrule(lr){3-6} \cmidrule(lr){7-9}
\textbf{Dataset} & \textbf{Model} 
& \textbf{WER(\%)}$\downarrow$ 
& \textbf{CER(\%)}$\downarrow$ 
& \textbf{SECS}$\uparrow$ 
& \textbf{$\Delta$PPG}$\downarrow$ 
& \textbf{NMOS}$\uparrow$ 
& \textbf{SMOS}$\uparrow$ 
& \textbf{ACT}$\uparrow$ \\
\midrule

\multirow{8}{*}{\begin{tabular}[c]{@{}l@{}}Demo samples\\ provided by\\ corresponding\\ baselines\end{tabular}} 
& MLVAE-ADV~\cite{melechovsky2024accent} 
& \textbf{14.30} & 11.10 & 0.74 & 0.64 
& $2.48\pm0.17$ & $3.03\pm0.17$ & $3.12\pm0.16$ \\

& Ours 
& 18.20 & \textbf{11.00} & \textbf{0.79} & \textbf{0.50} 
& $\mathbf{4.13\pm0.13}$ & $\mathbf{3.95\pm0.15}$ & $\mathbf{3.91\pm0.13}$ \\

\cmidrule{2-9}

& TTS-AC~\cite{chen2024transfer} 
& 15.28 & 10.88 & 0.76 & 0.57 
& $3.91\pm0.15$ & $3.78\pm0.18$ & $3.74\pm0.19$ \\

& Ours 
& \textbf{5.00} & \textbf{4.41} & \textbf{0.81} & \textbf{0.46} 
& $\mathbf{4.31\pm0.13}$ & $\mathbf{4.24\pm0.12}$ & $\mathbf{4.03\pm0.14}$ \\

\cmidrule{2-9}

& VITS-AC~\cite{nguyen2025improving} 
& \textbf{11.84} & \textbf{6.67} & 0.70 & 0.38 
& $3.93\pm0.11$ & $3.67\pm0.12$ & $\mathbf{3.97\pm0.11}$ \\

& Ours 
& 15.13 & 7.67 & \textbf{0.81} & \textbf{0.35} 
& $\mathbf{3.99\pm0.12}$ & $\mathbf{3.73\pm0.11}$ & $3.95\pm0.11$ \\

\cmidrule{2-9}

& VP-AC~\cite{jin2023voice} 
& 4.00 & 0.87 & 0.76 & 0.30 
& $4.11\pm0.09$ & $4.01\pm0.10$ & $\mathbf{4.17\pm0.11}$ \\

& Ours 
& \textbf{0.00} & \textbf{0.00} & \textbf{0.82} & \textbf{0.26} 
& $\mathbf{4.19\pm0.10}$ & $\mathbf{4.05\pm0.10}$ & $4.11\pm0.10$ \\

\bottomrule
\end{tabular*}
\end{table*}

\subsection{Evaluation Metrics}
Both objective and subjective metrics are employed to comprehensively assess conversion performances.

\subsubsection{Objective Metrics}
For `Reproducible' baselines, objective metrics are computed on the entire corpus of the L2-ARCTIC (ASI, ABA) and VCTK (P248) datasets (comprising 1131 utterances for ASI, 1129 utterances for ABA, and 376 utterances for P248, without applying any test split), while for `Demos' baselines, objective metrics are computed on the official demo samples:
\begin{itemize}
    \item \textbf{Intelligibility:} The content preservation performance is assessed via Word Error Rate (WER) and Character Error Rate (CER) of the converted speech, computed by comparing Whisper Large V3~\cite{radford2023robust} transcriptions against ground-truth text.
    
    \item \textbf{Speaker Preservation:} Speaker identity retention performance is quantified using Speaker Embedding Cosine Similarity (SECS) between source  and converted speech, with embeddings extracted via Resemblyzer~\cite{wan2018generalized} following~\cite{baas2023voice}. 
     
    \item \textbf{PPG Distance ($\Delta$PPG):} The accent fidelity of the converted speech is measured by computing the distance between high-fidelity neural PPGs~\cite{churchwell2024high} of converted speech and those of the parallel native references, i.e., $\Delta$PPG (speaker BDL from CMU-ARCTIC for L2-ARCTIC, and speaker P297 containing linguistically identical content for VCTK).
    \item \textbf{Real-Time Factor (RTF):} The inference efficiency of the model is evaluated via the Real-Time Factor (RTF), computed as the ratio of the total processing time required to generate the converted speech to the total duration of the resulting audio.
\end{itemize}
\subsubsection{Subjective Metrics}
Mean Opinion Score (MOS) tests are conducted following ITU-T P.800 recommendations. Twenty listeners rate samples on a 5-point scale. For `Reproducible' baselines, 10 utterances are randomly selected per test set; for `Demos' baselines, all available demos are evaluated.
\begin{itemize}
    \item \textbf{Naturalness (NMOS):} Measures the overall speech quality and fluency (1: Poor; 5: Excellent).
    
    \item \textbf{Similarity (SMOS):} Measures the resemblance of the converted speaker identity to the source speaker (1: Different speaker; 5: Identical speaker).
    
    \item \textbf{Accentedness (ACT):} Measures the degree of native-like pronunciation (1: Strong foreign accent; 5: Native-like).
\end{itemize}
\section{Experimental Results}

\subsection{Comparison with Baseline Models}

We compare the performances of the proposed model with those of various state-of-the-art approaches `Reproducible' models (see Table~\ref{tab:baseline1}) and `Demos' models (see Table~\ref{tab:baseline2}).

\subsubsection{Comparison with `Reproducible' Baselines}
Table~\ref{tab:baseline1} shows three main trends:

\begin{enumerate}
    \item[i)] Regardless of the evaluation metric (objective or subjective) considered, the proposed model consistently outperforms the previously proposed retrieval-based models, i.e., $k$NN-VC~\cite{baas2023voice} and $k$NN-AC~\cite{yixin2025zero}. Specifically, when compared with $k$NN-AC~\cite{yixin2025zero}, the reductions in WER and CER are 36.36\% and 38.69\%, respectively, and the improvement in SMOS is 13.86\%. These improvements are consistent with the intended effects of the phoneme-coverage-aware matching pool and the anisotropy-aware adaptive retrieval strategy.

    \item[ii)] When objective evaluation is considered, the proposed model achieves performance comparable to TokAN~\cite{bai2025accent} on most objective metrics, even though the training dataset adopted by the proposed model is much smaller than those adopted by TokAN. This result is largely attributed to the introduction of the phoneme coverage-based matching pool construction and the AA-$k$NN-based retrieval strategy. Furthermore, the superiority of the proposed model over TokAN on VCTK is much more prominent than that on the other datasets. The possible reason is that unlike L2-ARCTIC (ASI) or L2-ARCTIC (ABA), VCTK (P248) is not included in the training datasets (see Table~\ref{tab:baselines}) of TokAN~\cite{bai2025accent}. This result suggests that the proposed retrieval-based framework can generalize competitively to source speakers not covered by the training data. When subjective evaluation is considered, in 5 out of 9 metrics TokAN outperforms the proposed model. The possible reason is that the training set adopted by TokAN includes L2-ARCTIC; that is, it accesses the accent information at the training stage, while the proposed model does not use source-accent data during model training and performs conversion with a fixed target-accent matching pool at inference. Furthermore, the training of TokAN includes several steps, which increases the computational complexity.

    \item[iii)] Compared with Seed-VC~\cite{liu2024}, a diffusion-based voice conversion model, the proposed model achieves lower WER on all three evaluation sets and slightly higher NMOS scores. This indicates that although Seed-VC can generate natural speech, the proposed retrieval-based framework is more effective in preserving linguistic content for accent conversion by directly retrieving target-accent content features from a fixed matching pool.
\end{enumerate}

\subsubsection{Comparison with `Demos' of Baselines}
From Table~\ref{tab:baseline2}, it can be seen that:

\begin{enumerate}
    \item[i)] Regardless of the metric (objective or subjective) considered, the proposed model consistently outperforms TTS-AC~\cite{chen2024transfer} prominently.

    \item[ii)] When compared with MLVAE-ADV~\cite{melechovsky2024accent}, the proposed model achieves higher performance in terms of all metrics except for WER. In particular, although the WER is higher, the CER is slightly lower, suggesting that the remaining recognition errors are more concentrated at the word level rather than the character level.

    \item[iii)] It outperforms VP-AC~\cite{jin2023voice} in terms of all objective metrics and achieves equivalent subjective metrics to VP-AC~\cite{jin2023voice}, even though VP-AC includes L2-ARCTIC in the training dataset.

    \item[iv)] Compared with VITS-AC~\cite{nguyen2025improving}, the proposed model achieves better SECS and $\Delta$PPG but higher WER and CER, while the subjective scores remain comparable.
\end{enumerate}

Overall, these results show that the proposed model remains competitive under the fixed target-accent matching-pool setting, especially when compared with retrieval-based and AC-oriented baselines.

\subsection{Ablation Studies}
\label{sec:ablation}

Considering that the main contributions of this paper are the phoneme coverage-based matching pool construction, the AA-$k$NN retrieval strategy, and the progressive AdaIN-based decoder, the effectiveness of each of these key modules is studied in comparison with the conventional strategy. 

\subsubsection{Effectiveness of Matching Pool Construction Strategy}

\begin{figure}
\centering
\includegraphics[width=1\columnwidth]{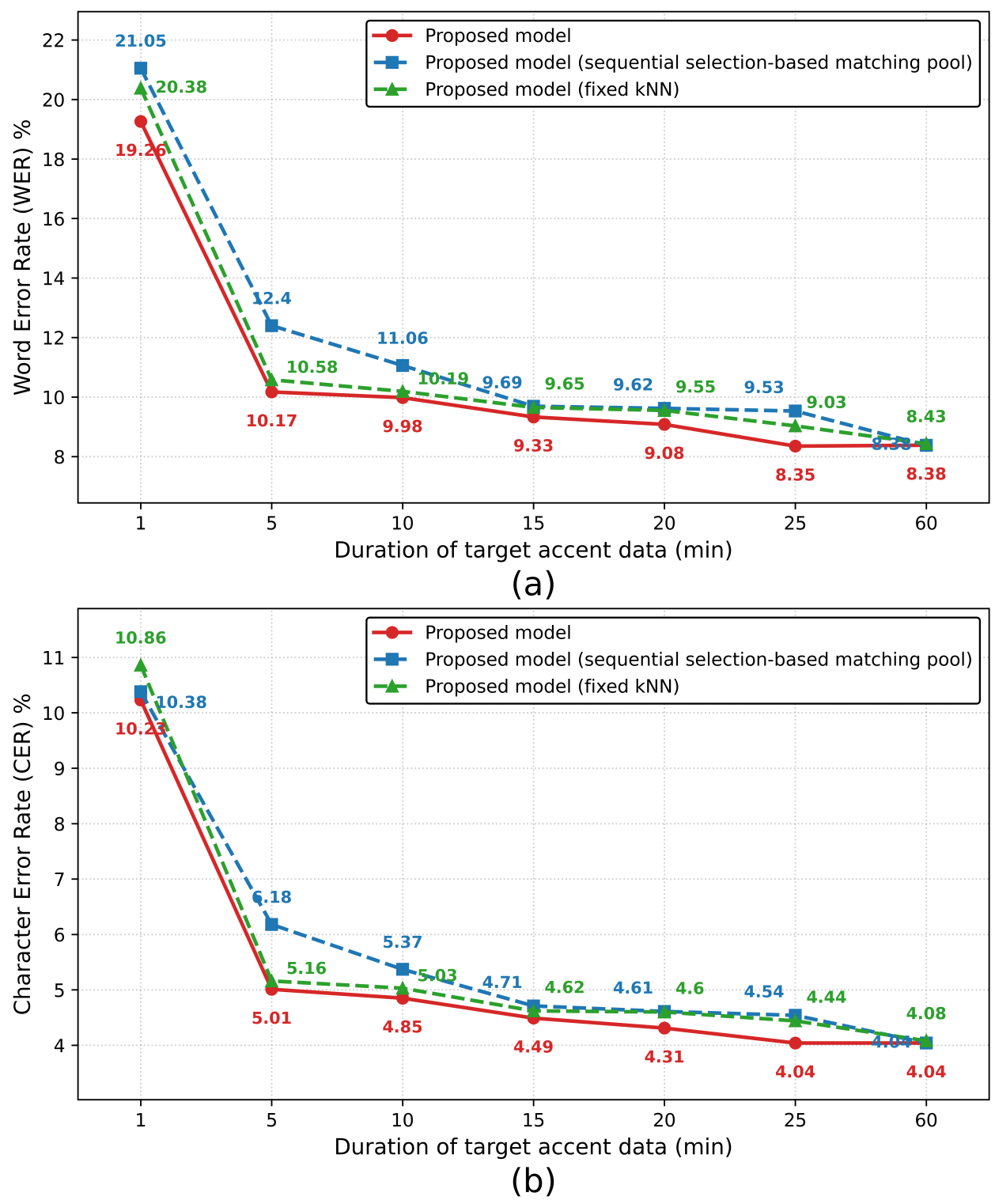} 
\caption{Superiority of the proposed matching pool construction and AA-$k$NN strategies over the sequential selection-based construction strategy and fixed $k$NN strategy.}
\label{fig:ablation_strategy}
\end{figure}

As detailed in Section 3.1, our method employs a composite quality function $Q(x_k)$, which integrates phoneme uniqueness $\mathcal{U}_k$, Shannon diversity index $H(x_k)$, category diversity $\mathcal{D}_k$, rare phoneme coverage $R_k$, and distribution uniformity $E_k$. To demonstrate the superiority of the proposed phoneme coverage-based matching pool construction strategy over the sequential selection-based matching pool construction, where speech samples are selected in their original chronological order, we compare the WER and CER achieved by these two methods under various target-accent data durations (1--60 minutes).

Fig.~\ref{fig:ablation_strategy} shows the WER and CER performance comparison results across different target-accent data durations. Focusing on the comparison between the proposed model and its obtained version using the sequential selection-based matching pool, it can be seen that:
\begin{enumerate}
    \item[i)] Under both evaluation metrics, the proposed model consistently outperforms its obtained version when the duration of target-accent data is less than or equal to 25~minutes, indicating that the proposed matching-pool construction strategy is more effective.
    \item[ii)] For the proposed model, the performance under the 25-minute duration is comparable to that under the 60-minute duration. In contrast, for its obtained version, the performance under the 25-minute duration is worse than that under the 60-minute duration. This indicates that the proposed matching-pool construction strategy can achieve comparable performance with a shorter duration of target-accent data.
\end{enumerate}

\begin{figure}
    \centering
    \includegraphics[width=\columnwidth]{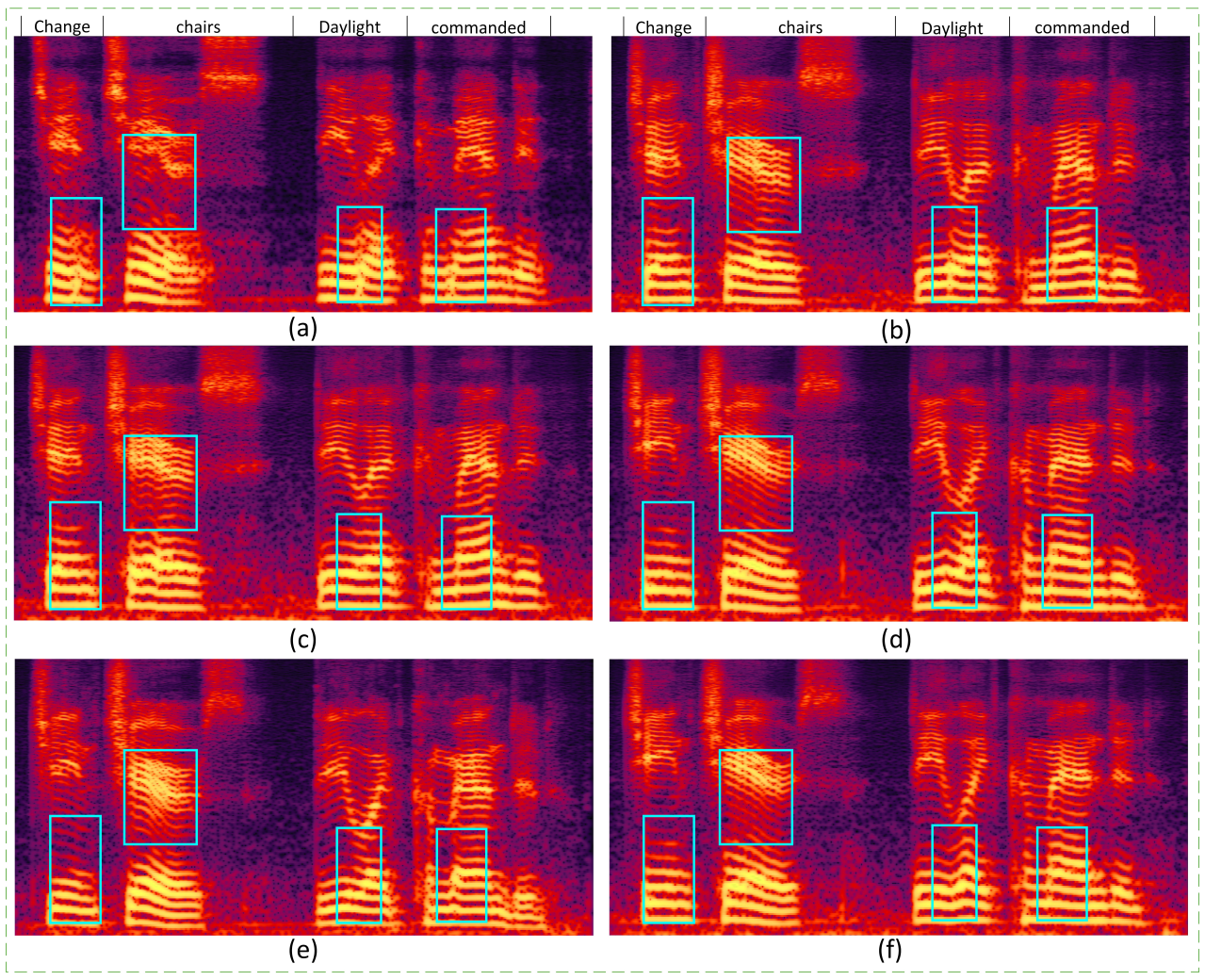}
    \caption{Mel-spectrograms generated by different model settings.
    (a) $k$NN-AC~\cite{yixin2025zero}, which uses sequential selection-based matching pool construction, fixed-$k$ $k$NN retrieval, and a dual-stage AdaIN decoder.
    (b) The proposed framework using sequential selection-based matching pool construction, fixed-$k$ $k$NN retrieval, and the progressive AdaIN decoder.
    (c) The proposed framework using sequential selection-based matching pool construction, AA-$k$NN retrieval, and the progressive AdaIN decoder.
    (d) The proposed framework using phoneme coverage-based matching pool construction, fixed-$k$ $k$NN retrieval, and the progressive AdaIN decoder.
    (e) The proposed framework using phoneme coverage-based matching pool construction, AA-$k$NN retrieval, and a dual-stage AdaIN decoder.
    (f) The proposed model, which combines phoneme coverage-based matching pool construction, AA-$k$NN retrieval, and the progressive AdaIN decoder.
    White boxes indicate representative regions used for qualitative comparison.}
    \label{fig:spectrogram_comparison}
\end{figure}

In addition, Fig.~\ref{fig:spectrogram_comparison} shows the Mel-spectrograms achieved by different models. By comparing the Mel-spectrogram achieved by the proposed model (Fig.~\ref{fig:spectrogram_comparison}(f)) and that by its obtained version (Fig.~\ref{fig:spectrogram_comparison}(c)), we can see that the formant structures appear indistinct in (c), likely due to deficiencies in the conventionally constructed matching pool. In contrast, Fig.~\ref{fig:spectrogram_comparison}(f) exhibits distinct formant trajectories and sharp phoneme transitions, indirectly demonstrating the superiority of the proposed matching pool construction strategy.

\subsubsection{Influence of the combination of Weights Assignment in Quality Evaluation Function}

\begin{table*}[!t]
\centering
\caption{Performance comparison under different combinations of the weights in the quality evaluation function in Eq.~(1)}
\label{tab:weight_ablation}
\tablefontsize
\setlength{\tabcolsep}{6pt}
\renewcommand{\arraystretch}{1.08}
\begin{tabular}{@{}llcccc@{}}
\toprule
\textbf{Configuration} 
& \textbf{Weights $(\lambda_1, \lambda_2, \lambda_3, \lambda_4, \lambda_5)$} 
& \textbf{WER(\%)$\downarrow$} 
& \textbf{CER(\%)$\downarrow$} 
& \textbf{SECS$\uparrow$} 
& \textbf{$\Delta$PPG$\downarrow$} \\
\midrule
Proposed setting
& (0.20, 0.30, 0.05, 0.40, 0.05) & \textbf{8.35} & \textbf{4.04} & \textbf{0.80} & \textbf{0.36} \\

Equal Weights 
& (0.20, 0.20, 0.20, 0.20, 0.20) & 8.69 & 4.26 & 0.79 & 0.37 \\

w/o Unique Phonemes
& (0.00, 0.40, 0.15, 0.30, 0.15) & 9.11 & 4.38 & 0.79 & 0.37 \\

w/o Diversity
& (0.30, 0.00, 0.10, 0.50, 0.10) & 9.05 & 4.41 & 0.78 & 0.38 \\

w/o Phoneme Categories
& (0.25, 0.35, 0.00, 0.30, 0.10) & 8.96 & 4.32 & 0.79 & 0.37 \\

w/o Rare Phonemes
& (0.35, 0.45, 0.10, 0.00, 0.10) & 9.15 & 4.47 & 0.79 & 0.37 \\

w/o Distribution Evenness
& (0.15, 0.35, 0.05, 0.45, 0.00) & 8.92 & 4.27 & \textbf{0.80} & 0.37 \\
\bottomrule
\end{tabular}
\end{table*}

To further validate the heuristic weighting strategy in the quality evaluation function $Q(x_k)$ defined in Eq.~(1), we conduct performance comparison among different combinations of its five components $\lambda_1 \sim \lambda_5$, of the unique phoneme count $\mathcal{U}_k$, the Shannon diversity index $H(x_k)$, the broad phoneme category count $\mathcal{D}_k$, the rare phoneme count $R_k$, and the distribution evenness $E_k$. In addition to the proposed model, we evaluate an equal-weight setting and five variants in which each component is individually removed. As shown in Table~\ref{tab:weight_ablation}, the proposed model achieves the best WER/CER of 8.35\%/4.04\%, demonstrating that the adopted weight assignment is effective for constructing a high-quality target-accent matching pool.

Compared with the proposed model, assigning equal weights to all components increases WER/CER to 8.69\%\slash 4.26\%, which suggests that the five phonetic criteria contribute unequally to matching-pool construction. Since natural speech usually exhibits an imbalanced phoneme distribution, treating all criteria equally may weaken the emphasis on retrieval-critical factors, particularly rare phoneme coverage and phoneme-level diversity.

When each component is removed individually, all variants show higher WER and CER compared with the proposed model. Removing the rare phoneme term $R_k$ causes the largest degradation, increasing WER/CER to 9.15\%/4.47\%, which confirms the importance of explicitly prioritizing low-frequency phonemes during matching-pool construction. Removing the unique phoneme count $\mathcal{U}_k$ also leads to a clear performance drop, with WER/CER increasing to 9.11\%/4.38\%, indicating that broad phoneme coverage is necessary for reliable retrieval. Similarly, removing the Shannon diversity term $H(x_k)$ increases WER/CER to 9.05\%/4.41\%, showing that phoneme-level diversity is important for selecting a balanced set of target-accent utterances.

The removal of $\mathcal{D}_k$ and $E_k$ results in relatively smaller but still noticeable degradation. Specifically, removing the broad phoneme category count $\mathcal{D}_k$ increases WER/CER to 8.96\%/4.32\%, while removing the distribution evenness term $E_k$ increases WER/CER to 8.92\%/4.27\%. These results indicate that $\mathcal{D}_k$ and $E_k$ act as complementary constraints: the former helps maintain coarse-grained coverage across major phoneme categories, while the latter discourages overly concentrated phoneme distributions.

In contrast, SECS and $\Delta$PPG show only minor variations across different weighting configurations. This is expected because the ablation only modifies the weighting strategy used for matching-pool sample selection, while the target-accent corpus and decoder architecture remain unchanged. Since all configurations construct the matching pool from the same American target-accent corpus and use the same speaker-content reconstruction module, the overall speaker similarity and accent-related distance remain relatively stable. Therefore, the proposed model mainly affects phoneme coverage and local retrieval accuracy through the weighting strategy in Eq.~(1), which is more directly reflected in WER and CER.

Overall, these results support the design rationale of Eq.~(1). The proposed model emphasizes rare phoneme coverage and phoneme-level diversity while retaining phoneme uniqueness, category coverage, and distribution evenness as complementary constraints. Although the coefficients used in the proposed model are not claimed to be globally optimal, the ablation results show that the proposed model is a robust empirical choice among several reasonable alternatives for constructing a balanced and phonetically representative matching pool.

\subsubsection{Effectiveness of Boundary Handling and Anisotropy Scaling}

Table~\ref{tab:boundary_ablation} evaluates the effects of boundary handling and anisotropy score correction through several controlled settings, including without anisotropy score correction, unit scaling, linear scaling, and the proposed square-root correction in Eq.~(8). As described in Section 3.2.1, frames near the beginning or end of an utterance have shorter receptive fields than interior frames, i.e., $w_j < 2\lfloor w/2 \rfloor + 1$. Therefore, we compare the proposed square-root scaling in Eq.~(8) with three boundary-related variants, including no correction, unit scaling, and linear scaling, to examine whether boundary handling and the specific scaling form affect retrieval performance.

First, removing the anisotropy score correction(i.e. anisotropy score correction) increases WER and CER from 8.35\%/4.04\% to 8.92\%/4.35\%, respectively. This result shows that directly using the uncorrected anisotropy score for boundary frames weakens local matching accuracy, because the shortened receptive field is not explicitly compensated. The variant with unit scaling, which replaces the correction factor in Eq.~(8) with $1$(i.e. 'Unit scale'), obtains similar WER/CER of 8.92\%/4.32\%. This is expected because unit scaling keeps the formulation of Eq.~(8) but removes the actual magnitude compensation, and therefore behaves similarly to the no-correction setting.

Second, we further ablate the square-root form of Eq.~(8) by replacing it with a linear scaling factor $(2\lfloor w/2 \rfloor + 1)/w_j$(i.e. 'Linear scale'). Compared without anisotropy score correction, linear scaling partially improves WER/CER to 8.77\%/4.24\%, indicating that compensating for the shortened receptive field is beneficial. However, it still performs worse than the proposed square-root correction. This suggests that a linear factor may over-correct the anisotropy scores of boundary frames, whereas the square-root factor provides a milder and more stable compensation.

While, SECS and $\Delta$PPG remain nearly unchanged under different correction settings, as indicates that the correction strategy mainly affects intelligibility-related metrics by improving local retrieval reliability in AA-$k$NN, while having limited influence on speaker similarity and phonetic-content consistency. Overall, these results verify both the necessity of anisotropy score correction and the effectiveness of the square-root scaling factor in Eq.~(8).

\begin{table*}[t]
\centering
\caption{Effectiveness of anisotropy score correction strategy in Eq. (8)}
\label{tab:boundary_ablation}
\footnotesize
\setlength{\tabcolsep}{12pt}
\renewcommand{\arraystretch}{1.08}
\begin{tabular}{@{}lcccc@{}}
\toprule
\textbf{Configuration} 
& \textbf{WER(\%)$\downarrow$} 
& \textbf{CER(\%)$\downarrow$} 
& \textbf{SECS$\uparrow$} 
& \textbf{$\Delta$PPG$\downarrow$} \\
\midrule
\textbf{Proposed Model} 
& \textbf{8.35} & \textbf{4.04} & \textbf{0.80} & \textbf{0.36} \\
w/o anisotropy score correction 
& 8.92 & 4.35 & 0.79 & 0.37 \\
Unit scale($\sqrt{(2\lfloor w/2 \rfloor + 1)/w_j}\to 1$)
& 8.92 & 4.32 & 0.79 & 0.37 \\
Linear scale($\sqrt{(2\lfloor w/2 \rfloor + 1)/w_j}\to (2\lfloor w/2 \rfloor + 1)/w_j$)
& 8.77 & 4.24 & 0.79 & 0.37 \\
\bottomrule
\end{tabular}
\end{table*}

\subsubsection{Effectiveness of AA-\texorpdfstring{$k$}{k}NN}



To demonstrate the effectiveness of the proposed AA-$k$NN-based matching strategy, which dynamically adjusts $k$ within $[2, 6]$, we compare the proposed model with its modified version, in which AA-$k$NN is replaced by conventional fixed-$k$ $k$NN.
As shown in Fig.~\ref{fig:ablation_strategy}, the proposed AA-$k$NN strategy consistently achieves lower WER and CER than the fixed-$k$ counterpart across different target-accent data durations, 
as indicates that adapting the neighborhood size according to local feature-space geometry is more effective than using a fixed number of neighbors. 
Moreover, the proposed model with a 25-minute target-accent matching pool achieves performance comparable to the fixed-$k$ version with a 60-minute pool, suggesting that AA-$k$NN uses the constructed matching pool more efficiently.
The qualitative comparison in Fig.~\ref{fig:spectrogram_comparison} further supports this conclusion. 
By comparing Fig.~\ref{fig:spectrogram_comparison}(d) and Fig.~\ref{fig:spectrogram_comparison}(f), it can be observed that the fixed-$k$ strategy produces more spectral smearing around rapid phonetic transitions. 
This is because a fixed neighborhood size is less capable of accurately matching features in highly anisotropic regions, where more precise retrieval is required. 
In contrast, AA-$k$NN dynamically reduces $k$ in high-anisotropy regions to preserve transient details, while using a larger $k$ in low-anisotropy regions to maintain stable reconstruction.

\subsubsection{Effectiveness of Progressive AdaIN-based Fine-grained Decoder}

To quantitatively evaluate the contribution of the progressive AdaIN-based decoder, we compare the proposed model with its modified version without the progressive AdaIN architecture.

\begin{table}[t]
\centering
\caption{Ablation of the progressive AdaIN decoder.}
\label{tab:adain_ablation}
\footnotesize
\renewcommand{\arraystretch}{1.08}
\begin{tabular*}{\columnwidth}{@{\extracolsep{\fill}}lcccc@{}}
\toprule
\textbf{Configuration} 
& \textbf{WER$\downarrow$} 
& \textbf{CER$\downarrow$} 
& \textbf{SECS$\uparrow$} 
& \textbf{$\Delta$PPG$\downarrow$} \\
\midrule
\textbf{Proposed Model} & \textbf{8.35} & \textbf{4.04} & \textbf{0.80} & \textbf{0.36} \\
w/o progressive AdaIN & 9.06 & 4.42 & 0.77 & 0.39 \\
\bottomrule
\end{tabular*}
\end{table}

As shown in Table~\ref{tab:adain_ablation}, removing the progressive AdaIN decoder increases the WER and CER from 8.35\%/4.04\% to 9.06\%/4.42\%, respectively. 
Meanwhile, SECS decreases from 0.80 to 0.77, and $\Delta$PPG increases from 0.36 to 0.39. 
These results indicate that the progressive AdaIN decoder contributes not only to intelligibility improvement, but also to more stable source speaker identity preservation and a smaller accent-related PPG distance.

The qualitative comparison in Fig.~\ref{fig:spectrogram_comparison} further supports this conclusion. 
In particular, comparing Fig.~\ref{fig:spectrogram_comparison}(f) with Fig.~\ref{fig:spectrogram_comparison}(e) reveals that the proposed model produces more distinct and continuous high-frequency harmonic patterns. 
This advantage stems from the progressive AdaIN design, which performs style injection at multiple temporal resolutions and couples speaker-content integration with temporal up-sampling. 
As a result, the decoder can better preserve fine-grained spectral details and reduce harmonic discontinuities, thereby improving the overall spectral continuity and intelligibility of the converted speech.

\subsection{Extended Evaluations on Cross-Gender Robustness and Bidirectional Generalization}

\begin{table*}[!t]
\centering
\caption{Performance comparison under other-to-American and American-to-other conversion scenarios across different genders and accents}
\label{tab:performance_comparison}
\tablefontsize
\setlength{\tabcolsep}{5pt}
\renewcommand{\arraystretch}{1.08}
\begin{tabular}{@{}ccccccc@{}}
\toprule
\textbf{Direction}
& \textbf{Source} 
& \textbf{Target Pool} 
& \textbf{WER(\%)$\downarrow$} 
& \textbf{CER(\%)$\downarrow$} 
& \textbf{SECS$\uparrow$} 
& \textbf{$\Delta$PPG$\downarrow$} \\
\midrule

\multirow{6}{*}{\begin{tabular}[c]{@{}c@{}}Other-to-\\American\end{tabular}}
& \multirow{2}{*}{ASI(M)}  
& BDL(M) & \textbf{8.35} & \textbf{4.04} & \textbf{0.80} & 0.36 \\
& & SLT(F) & 9.21 & 4.58 & 0.78 & \textbf{0.33} \\
\cmidrule(lr){2-7}

& \multirow{2}{*}{ABA(M)}  
& BDL(M) & \textbf{15.72} & \textbf{7.31} & \textbf{0.82} & 0.51 \\
& & SLT(F) & 16.56 & 8.01 & 0.80 & \textbf{0.46} \\
\cmidrule(lr){2-7}

& \multirow{2}{*}{P248(F)} 
& BDL(M) & 5.86 & 2.37 & \textbf{0.78} & \textbf{0.30} \\
& & SLT(F) & \textbf{5.69} & \textbf{2.34} & 0.77 & 0.31 \\

\midrule

\multirow{4}{*}{\begin{tabular}[c]{@{}c@{}}American-to-\\Other\end{tabular}}
& \multirow{2}{*}{BDL(M)}  
& ASI(M) & \textbf{4.67} & \textbf{2.30} & \textbf{0.83} & \textbf{0.43} \\
& & ABA(M) & 5.17 & 2.47 & \textbf{0.83} & 0.55 \\
\cmidrule(lr){2-7}

& \multirow{2}{*}{SLT(F)}  
& ASI(M) & \textbf{4.68} & \textbf{2.00} & 0.85 & \textbf{0.44} \\
& & ABA(M) & 7.70 & 4.32 & \textbf{0.86} & 0.53 \\
\bottomrule
\end{tabular}

\smallskip
\parbox{0.9\textwidth}{\tablefontsize \textit{Note:} M stands for male and F stands for female. Bold indicates the best performance in each group.}
\end{table*}

To further evaluate the robustness and bidirectional applicability of the proposed retrieval-based framework, we conduct extended evaluations on both other-accent-to-American and American-to-other-accent conversion settings. 
Specifically, the first setting evaluates the robustness of converting other-accented English speech to American-accented English using target-accent matching pools constructed from speakers of different genders, while the second setting examines the reverse direction, i.e., converting American-accented English speech to other non-American accents. 
The objective results are summarized in Table~\ref{tab:performance_comparison}.

\textbf{1) Performance under Other-to-American Accent Conversion Scenario:} 
As shown in the top half of Table~\ref{tab:performance_comparison}, when converting other-accented source speech (ASI, ABA, and P248) to an American accent, the performance remains stable when the American target-accent matching pool is constructed from the male speaker BDL or the female speaker SLT. 
For instance, the WER/CER of ASI-to-American conversion changes only from 8.35\%/4.04\% to 9.21\%/4.58\%, while the SECS remains close. 
Similar trends are also observed for ABA and P248. 
These results demonstrate that the proposed framework is robust to the gender of the speaker used to construct the American target-accent matching pool and can effectively preserve the source speaker's identity under cross-gender target-accent matching-pool conditions.

\textbf{2) Performance under American-to-Other Accent Conversion Scenario:} 
The bottom half of Table~\ref{tab:performance_comparison} evaluates the reverse conversion direction, i.e., converting American-accented English speech from BDL and SLT to other accents represented by ASI and ABA. 
The converted speech achieves relatively low WER and CER, e.g., 4.67\%/2.30\% for BDL-to-ASI conversion, together with consistently high SECS values ranging from 0.83 to 0.86.
This suggests that linguistic content and source speaker identity are well preserved in the American-to-other-accent conversion direction.

However, the $\Delta$PPG values in American-to-other-accent conversion are generally higher than those in the other-accent-to-American setting, indicating that target-accent realization is more challenging in this reverse direction. 
This phenomenon may be explained by two factors. 
First, the American-accented source speakers provide clear and highly intelligible pronunciations, which helps the content encoder extract stable linguistic representations and contributes to the low WER and CER. 
Second, since the AdaIN-based decoder and the HiFi-GAN vocoder are trained on American English speech, the waveform generation process may introduce an American-English acoustic prior. 
As a result, although the retrieved features are obtained from the non-American target pool, some fine-grained pronunciation characteristics of the synthesized speech may still be biased toward American English. 
This explains why the reverse conversion maintains high intelligibility and speaker similarity, while showing a relatively larger $\Delta$PPG distance to the non-American reference.

\subsection{Discussions}
Considering that the receptive field $w$ and the adaptive $k$ range influence the behavior of the AA-$k$NN-based matching strategy, we analyze the effects of these two hyperparameters within practical ranges.

\subsubsection{Influence of Receptive Field \texorpdfstring{$w$}{w}}
\begin{figure}
\centering
\includegraphics[width=\columnwidth]{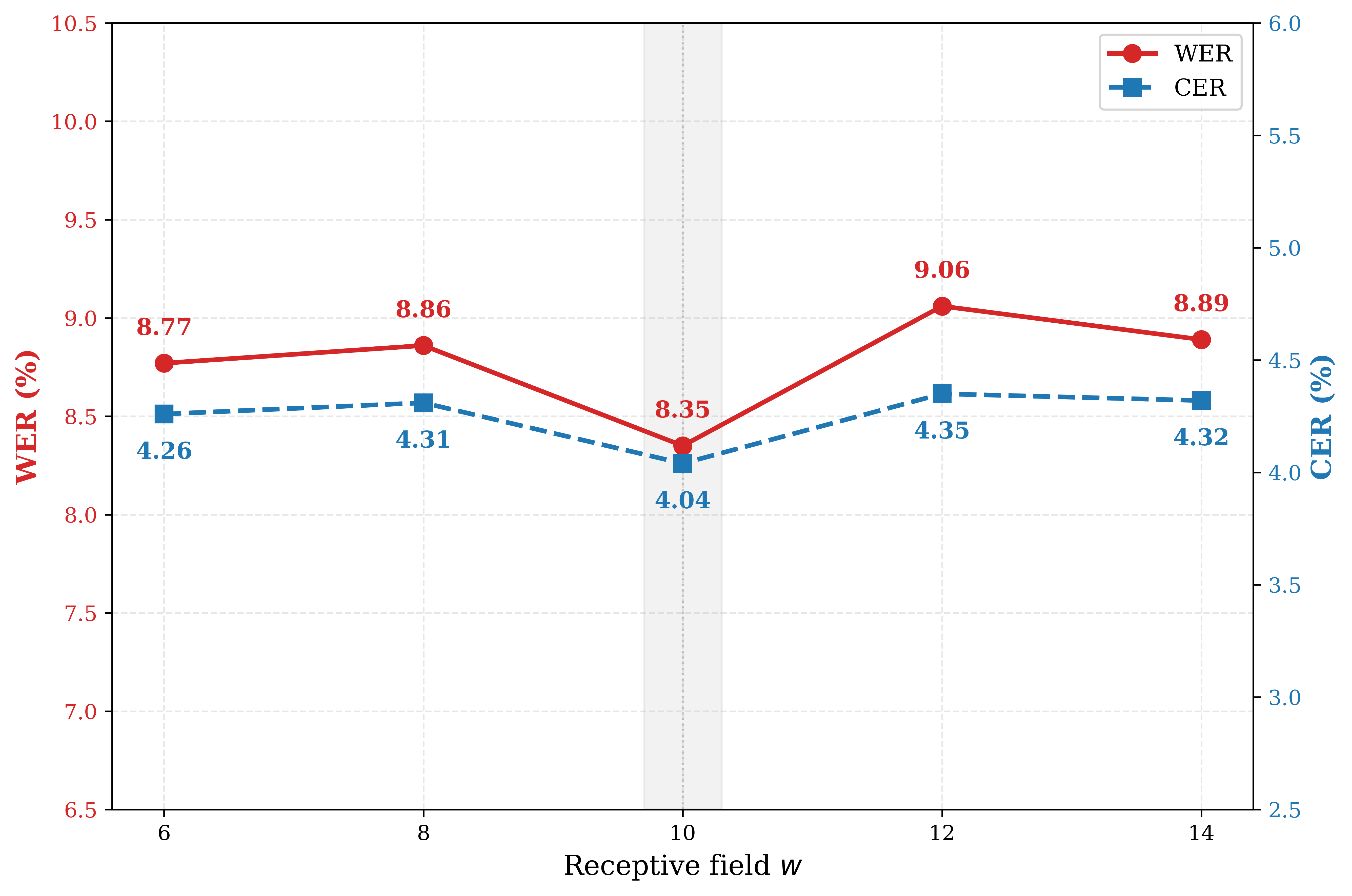}
\caption{Effect of receptive field \texorpdfstring{$w$}{w}. The U-shaped curve indicates that $w=10$ achieves the optimal balance between syllable-level coherence and statistical estimation stability.}
\label{fig:w_sensitivity}
\end{figure}
Considering that the receptive field influences the precision of spectral-variation estimation, the performance of the proposed model under different $w$ values, i.e., $6, 8, 10, 12, 14$, is tested. The experimental results shown in Fig.~\ref{fig:w_sensitivity} indicate that excessively large or small $w$ values degrade performance. The possible reasons are as follows. An excessively large receptive field may span multiple phonemes or syllables, which leads to higher anisotropy and a smaller neighborhood size. This will cause fluctuations in long vowels. In contrast, an excessively small receptive field may span only part of a phoneme or syllable, which leads to lower anisotropy and a larger neighborhood size. This will cause excessive smoothing of short consonants. Both cases affect the quality of the converted speech. Therefore, $w$ is set to a moderate value of 10. The performance varies smoothly across different receptive-field sizes rather than showing abrupt degradation around the selected value, indicating that the method remains stable within a reasonable neighborhood of this setting.

\subsubsection{Influence of Selection Range of \texorpdfstring{$k$}{k} }

AA-$k$NN maps the anisotropy percentile $\rho_q$ to a dynamic $k$ within $[k_{\min}, k_{\max}]$. We evaluate the performance of the proposed model under three configurations: [1, 7], [2, 6], and [3, 5]. As shown in Table~\ref{tab:k_range_ablation}, the range [2, 6] yields the best performance, which can be explained through bias-variance analysis:
\begin{itemize}
    \item \textbf{High Variance Risk with [1, 7]:} 
    Allowing $k=1$ makes the matching highly susceptible to idiosyncratic noise in the matching pool. A single outlier in a high-anisotropy region can dominate the output, potentially introducing audible artifacts and increasing both WER and CER.
    
    \item \textbf{High Bias Risk with [3, 5]:} 
    Restricting the range limits adaptability in two aspects. The lower bound $k_{\min}=3$ prevents precise matching along highly directional manifolds where $k=2$ would be optimal. Meanwhile, the upper bound $k_{\max}=5$ provides insufficient smoothing for isotropic regions, where larger $k$ values help suppress noise through averaging. This constrained range therefore increases both WER and CER relative to the proposed setting.
    
    \item \textbf{Optimal Trade-off with [2, 6]:} 
    This range effectively balances robustness against outliers ($k_{\min}=2$) with flexibility to adapt to diverse geometric structures ($k_{\max}=6$). The lower bound provides basic noise rejection while avoiding single-point sensitivity, and the upper bound offers adequate smoothing capacity without excessive averaging that would blur fine acoustic distinctions. As a result, this setting achieves the lowest WER and CER.
\end{itemize}
It is also observed that SECS and $\Delta$PPG remain nearly unchanged across different $k$ ranges (as shown in Table~\ref{tab:k_range_ablation}). 
This indicates that the adaptive $k$ range mainly affects local matching precision and intelligibility, as reflected by WER and CER, while speaker identity preservation and the overall accent-related distance are less sensitive to moderate changes in the $k$ range. These results show that $[2,6]$ provides the best trade-off and that the proposed strategy is robust within a practical adaptive-$k$ range.

\begin{table}[t]
\centering
\caption{Ablation of the adaptive $k$ range.}
\label{tab:k_range_ablation}
\footnotesize
\setlength{\tabcolsep}{4.5pt}
\renewcommand{\arraystretch}{1.08}
\begin{tabular*}{\columnwidth}{@{\extracolsep{\fill}}lcccc@{}}
\toprule
\textbf{Range} 
& \textbf{WER$\downarrow$} 
& \textbf{CER$\downarrow$} 
& \textbf{SECS$\uparrow$} 
& \textbf{$\Delta$PPG$\downarrow$} \\
\midrule
\textbf{$[2,6]$ Ours } 
& \textbf{8.35} & \textbf{4.04} & \textbf{0.80} & \textbf{0.36} \\
$[1,7]$ 
& 9.32 & 4.47 & 0.79 & \textbf{0.36} \\
$[3,5]$ 
& 8.90 & 4.31 & 0.79 & \textbf{0.36} \\
\bottomrule
\end{tabular*}
\end{table}

\section{Conclusion}

This paper presented AA-$k$NN-AC, a retrieval-based framework for accent conversion under a fixed target-accent matching-pool setting. The proposed framework consists of phoneme-coverage-aware matching-pool construction, anisotropy-aware adaptive $k$NN retrieval, and a progressive AdaIN-based decoder. Experiments on L2-ARCTIC and VCTK show that AA-$k$NN-AC improves over retrieval-based baselines in intelligibility, speaker similarity, and naturalness, and remains competitive with representative generative systems. Ablation studies further show that the proposed matching-pool construction, adaptive retrieval strategy, boundary correction, and progressive decoder each contribute to the final performance. Future work will focus on extending this paradigm to cross-lingual scenarios and developing low-latency streaming conversion for real-time applications.


\bibliographystyle{cas-model2-names}
\bibliography{references}

@INPROCEEDINGS{melechovsky2024accent,
  author={Melechovsky, Jan and Mehrish, Ambuj and Sisman, Berrak and Herremans, Dorien},
  booktitle={IEEE Region 10 Conference (TENCON)},
  title={Accent Conversion in Text-to-Speech Using Multi-Level VAE and Adversarial Training},
  year={2024},
  pages={473-476},
  doi={10.1109/TENCON61640.2024.10902878}
}

@ARTICLE{aryal2015reduction,
  author={Aryal, Sandesh and Gutierrez-Osuna, Ricardo},
  journal={The Journal of the Acoustical Society of America},
  title={Reduction of non-native accents through statistical parametric articulatory synthesis},
  year={2015},
  volume={137},
  number={1},
  pages={433-446},
  doi={10.1121/1.4904701}
}

@INPROCEEDINGS{zhao2019accent,
  author={Zhao, Guanlong and Sonsaat, Sinem and Levis, John and Chukharev-Hudilainen, Evgeny and Gutierrez-Osuna, Ricardo},
  booktitle={IEEE International Conference on Acoustics, Speech and Signal Processing (ICASSP)},
  title={Accent Conversion Using Phonetic Posteriorgrams},
  year={2018},
  pages={5314-5318},
  doi={10.1109/ICASSP.2018.8462258}
}

@INPROCEEDINGS{zhao2019using,
  author={Zhao, Guanlong and Ding, Shaojin and Gutierrez-Osuna, Ricardo},
  booktitle={INTERSPEECH 2019},
  title={Foreign Accent Conversion by Synthesizing Speech from Phonetic Posteriorgrams},
  year={2019},
  pages={2843-2847},
  doi={10.21437/Interspeech.2019-1778}
}

@INPROCEEDINGS{huang2024usd,
  author={Huang, Jen-Hung and Lee, Wei-Tsung and Wu, Chung-Hsien},
  booktitle={INTERSPEECH 2024},
  title={{USD-AC: Unsupervised Speech Disentanglement for Accent Conversion}},
  year={2024},
  pages={4388-4392},
  doi={10.21437/Interspeech.2024-664}
}

@INPROCEEDINGS{wang2021accent,
  author={Wang, Zhichao and Ge, Wenshuo and Wang, Xiong and Yang, Shan and Gan, Wendong and Chen, Haitao and Li, Hai and Xie, Lei and Li, Xiulin},
  booktitle={12th International Symposium on Chinese Spoken Language Processing (ISCSLP)},
  title={Accent and Speaker Disentanglement in Many-to-many Voice Conversion},
  year={2021},
  pages={1-5},
  doi={10.1109/ISCSLP49672.2021.9362120}
}

@ARTICLE{zhao2021converting,
  author={Zhao, Guanlong and Ding, Shaojin and Gutierrez-Osuna, Ricardo},
  journal={IEEE/ACM Transactions on Audio, Speech, and Language Processing},
  title={Converting Foreign Accent Speech Without a Reference},
  year={2021},
  volume={29},
  pages={2367-2381},
  doi={10.1109/TASLP.2021.3060813}
}

@INPROCEEDINGS{baas2023voice,
  author={Baas, Matthew and van Niekerk, Benjamin and Kamper, Herman},
  booktitle={INTERSPEECH 2023},
  title={Voice Conversion With Just Nearest Neighbors},
  year={2023},
  pages={2053-2057},
  doi={10.21437/Interspeech.2023-419}
}

@INPROCEEDINGS{bai2025accent,
  author={Bai, Qibing and Inoue, Sho and Wang, Shuai and Jiang, Zhongjie and Wang, Yannan and Li, Haizhou},
  booktitle={INTERSPEECH 2025},
  title={{Accent Normalization Using Self-Supervised Discrete Tokens with Non-Parallel Data}},
  year={2025},
  pages={1618-1622},
  doi={10.21437/Interspeech.2025-2520}
}

@INPROCEEDINGS{nguyen2025improving,
  author={Nguyen, Tuan Nam and Akti, Seymanur and Pham, Ngoc Quan and Waibel, Alexander},
  booktitle={IEEE International Conference on Acoustics, Speech and Signal Processing (ICASSP)},
  title={Improving Pronunciation and Accent Conversion through Knowledge Distillation And Synthetic Ground-Truth from Native TTS},
  year={2025},
  pages={1-5},
  doi={10.1109/ICASSP49660.2025.10890229}
}

@INPROCEEDINGS{levinson2020analysis,
  author={Levinson, Jake and Esteves, Carlos and Chen, Kefan and Snavely, Noah and Kanazawa, Angjoo and Rostamizadeh, Afshin and Makadia, Ameesh},
  booktitle={Advances in Neural Information Processing Systems},
  title={An Analysis of SVD for Deep Rotation Estimation},
  year={2020},
  volume={33},
  pages={22554-22565}
}

@ARTICLE{klema1980singular,
  author={Klema, V. and Laub, A.},
  journal={IEEE Transactions on Automatic Control},
  title={The singular value decomposition: Its computation and some applications},
  year={1980},
  volume={25},
  number={2},
  pages={164-176},
  doi={10.1109/TAC.1980.1102314}
}

@INPROCEEDINGS{jin2023voice,
  author={Jin, Mumin and Serai, Prashant and Wu, Jilong and Tjandra, Andros and Manohar, Vimal and He, Qing},
  booktitle={IEEE International Conference on Acoustics, Speech and Signal Processing (ICASSP)},
  title={Voice-Preserving Zero-Shot Multiple Accent Conversion},
  year={2023},
  pages={1-5},
  doi={10.1109/ICASSP49357.2023.10094737}
}

@ARTICLE{yixin2025zero,
  author={Luo, Yixin and Chen, Ning and Xue, Yuhang and Xiao, Yangyang},
  journal={Journal of East China University of Science and Technology},
  title={Zero-Shot Accent Conversion Model Based on the kNN Regression of Content Features},
  year={2025},
  doi={10.14135/j.cnki.1006-3080.20241009001}
}

@ARTICLE{ding2022accentron,
  author={Ding, Shaojin and Zhao, Guanlong and Gutierrez-Osuna, Ricardo},
  journal={Computer Speech \& Language},
  title={Accentron: Foreign accent conversion to arbitrary non-native speakers using zero-shot learning},
  year={2022},
  volume={72},
  pages={101302},
  doi={10.1016/j.csl.2021.101302}
}

@ARTICLE{zhou2023tts,
  author={Zhou, Yi and Wu, Zhizheng and Zhang, Mingyang and Tian, Xiaohai and Li, Haizhou},
  journal={IEEE Signal Processing Letters},
  title={TTS-Guided Training for Accent Conversion Without Parallel Data},
  year={2023},
  volume={30},
  pages={533-537},
  doi={10.1109/LSP.2023.3270079}
}

@ARTICLE{zhou2024accented,
  author={Zhou, Xuehao and Zhang, Mingyang and Zhou, Yi and Wu, Zhizheng and Li, Haizhou},
  journal={IEEE/ACM Transactions on Audio, Speech, and Language Processing},
  title={Accented Text-to-Speech Synthesis With Limited Data},
  year={2024},
  volume={32},
  pages={1699-1711},
  doi={10.1109/TASLP.2024.3363414}
}

@INPROCEEDINGS{baevski2020wav2vec,
  author={Baevski, Alexei and Zhou, Yuhao and Mohamed, Abdelrahman and Auli, Michael},
  booktitle={Advances in Neural Information Processing Systems},
  title={wav2vec 2.0: A Framework for Self-Supervised Learning of Speech Representations},
  year={2020},
  volume={33},
  pages={12449-12460}
}

@INPROCEEDINGS{chen2024transfer,
  author={Chen, Xi and Pei, Jiakun and Xue, Liumeng and Zhang, Mingyang},
  booktitle={IEEE International Conference on Acoustics, Speech and Signal Processing (ICASSP)},
  title={Transfer the Linguistic Representations from TTS to Accent Conversion with Non-Parallel Data},
  year={2024},
  pages={12501-12505},
  doi={10.1109/ICASSP48485.2024.10447205}
}

@INPROCEEDINGS{zhong2025accentbox,
  author={Zhong, Jinzuomu and Richmond, Korin and Su, Zhiba and Sun, Siqi},
  booktitle={IEEE International Conference on Acoustics, Speech and Signal Processing (ICASSP)},
  title={AccentBox: Towards High-Fidelity Zero-Shot Accent Generation},
  year={2025},
  pages={1-5},
  doi={10.1109/ICASSP49660.2025.10888332}
}

@INPROCEEDINGS{panayotov2015librispeech,
  author={Panayotov, Vassil and Chen, Guoguo and Povey, Daniel and Khudanpur, Sanjeev},
  booktitle={IEEE International Conference on Acoustics, Speech and Signal Processing (ICASSP)},
  title={Librispeech: An ASR corpus based on public domain audio books},
  year={2015},
  pages={5206-5210},
  doi={10.1109/ICASSP.2015.7178964}
}

@INPROCEEDINGS{kominek2004cmu,
  author={Kominek, John and Black, Alan W.},
  booktitle={5th ISCA Workshop on Speech Synthesis (SSW 5)},
  title={The CMU Arctic speech databases},
  year={2004},
  pages={223-224}
}

@INPROCEEDINGS{zhao2018l2,
  author={Zhao, Guanlong and Sonsaat, Sinem and Silpachai, Alif and Lucic, Ivana and Chukharev-Hudilainen, Evgeny and Levis, John and Gutierrez-Osuna, Ricardo},
  booktitle={INTERSPEECH 2018},
  title={L2-ARCTIC: A Non-native English Speech Corpus},
  year={2018},
  pages={2783-2787},
  doi={10.21437/Interspeech.2018-1110}
}

@INPROCEEDINGS{radford2023robust,
  author={Radford, Alec and Kim, Jong Wook and Xu, Tao and Brockman, Greg and McLeavey, Christine and Sutskever, Ilya},
  booktitle={International Conference on Machine Learning (ICML)},
  title={Robust Speech Recognition via Large-Scale Weak Supervision},
  year={2023},
  volume={202},
  pages={28492-28518}
}

@INPROCEEDINGS{wan2018generalized,
  author={Wan, Li and Wang, Quan and Papir, Alan and Moreno, Ignacio Lopez},
  booktitle={IEEE International Conference on Acoustics, Speech and Signal Processing (ICASSP)},
  title={Generalized End-to-End Loss for Speaker Verification},
  year={2018},
  pages={4879-4883},
  doi={10.1109/ICASSP.2018.8462665}
}

@INPROCEEDINGS{churchwell2024high,
  author={Churchwell, Cameron and Morrison, Max and Pardo, Bryan},
  booktitle={IEEE International Conference on Acoustics, Speech, and Signal Processing Workshops (ICASSPW)},
  title={High-Fidelity Neural Phonetic Posteriorgrams},
  year={2024},
  pages={823-827},
  doi={10.1109/ICASSPW62465.2024.10669905}
}

@ARTICLE{chen2022wavlm,
  author={Chen, Sanyuan and Wang, Chengyi and Chen, Zhengyang and Wu, Yu and Liu, Shujie and Chen, Zhuo and Li, Jinyu and Kanda, Naoyuki and Yoshioka, Takuya and Xiao, Xiong and others},
  journal={IEEE Journal of Selected Topics in Signal Processing},
  title={{WavLM}: Large-Scale Self-Supervised Pre-Training for Full Stack Speech Processing},
  year={2022},
  volume={16},
  number={6},
  pages={1505-1518}
}

@INPROCEEDINGS{qian2019autovc,
  author={Qian, Kaizhi and Zhang, Yang and Chang, Shiyu and Yang, Xuesong and Hasegawa-Johnson, Mark},
  booktitle={36th International Conference on Machine Learning (ICML)},
  title={{AutoVC}: Zero-Shot Voice Style Transfer with Only Autoencoder Loss},
  year={2019},
  volume={97},
  pages={5210-5219}
}

@INPROCEEDINGS{chan2022speechsplit,
  author={Ho Chan, Chak and Qian, Kaizhi and Zhang, Yang and Hasegawa-Johnson, Mark},
  booktitle={IEEE International Conference on Acoustics, Speech and Signal Processing (ICASSP)},
  title={SpeechSplit2.0: Unsupervised Speech Disentanglement for Voice Conversion without Tuning Autoencoder Bottlenecks},
  year={2022},
  pages={6332-6336},
  doi={10.1109/ICASSP43922.2022.9747763}
}

@INPROCEEDINGS{yi2020vcc,
  author={Yi, Zhao and Huang, Wen-Chin and Tian, Xiaohai and Yamagishi, Junichi and Das, Rohan Kumar and Kinnunen, Tomi and Ling, Zhen-Hua and Toda, Tomoki},
  booktitle={Joint Workshop for the Blizzard Challenge and Voice Conversion Challenge 2020},
  title={Voice Conversion Challenge 2020 -- Intra-lingual semi-parallel and cross-lingual voice conversion},
  year={2020},
  pages={80-98},
  doi={10.21437/VCCBC.2020-14}
}

@INPROCEEDINGS{lorenzotrueba2018vcc,
  author={Lorenzo-Trueba, Jaime and Yamagishi, Junichi and Toda, Tomoki and Saito, Daisuke and Villavicencio, Fernando and Kinnunen, Tomi and Ling, Zhenhua},
  booktitle={The Speaker and Language Recognition Workshop (Odyssey 2018)},
  title={The Voice Conversion Challenge 2018: Promoting Development of Parallel and Nonparallel Methods},
  year={2018},
  pages={195-202},
  doi={10.21437/Odyssey.2018-28}
}

@INPROCEEDINGS{ma2024accentvits,
  author={Ma, Linhan and Zhang, Yongmao and Zhu, Xinfa and Lei, Yi and Ning, Ziqian and Zhu, Pengcheng and Xie, Lei},
  booktitle={Man-Machine Speech Communication},
  title={Accent-VITS: Accent Transfer for End-to-End TTS},
  year={2024},
  pages={203-214}
}

@INPROCEEDINGS{liu2020end,
  author={Liu, Songxiang and Wang, Disong and Cao, Yuewen and Sun, Lifa and Wu, Xixin and Kang, Shiyin and Wu, Zhiyong and Liu, Xunying and Su, Dan and Yu, Dong and Meng, Helen},
  booktitle={IEEE International Conference on Acoustics, Speech and Signal Processing (ICASSP)},
  title={End-To-End Accent Conversion Without Using Native Utterances},
  year={2020},
  pages={6289-6293},
  doi={10.1109/ICASSP40776.2020.9053797}
}

@INPROCEEDINGS{wu2020vqvc,
  author={Wu, Da-Yi and Chen, Yen-Hao and Lee, Hung-yi},
  booktitle={INTERSPEECH 2020},
  title={VQVC+: One-Shot Voice Conversion by Vector Quantization and U-Net Architecture},
  year={2020},
  pages={4691-4695},
  doi={10.21437/Interspeech.2020-1443}
}

@ARTICLE{zhou2023optimization,
  author={Zhou, Yi and Wu, Zhizheng and Tian, Xiaohai and Li, Haizhou},
  journal={IEEE/ACM Transactions on Audio, Speech, and Language Processing},
  title={Optimization of Cross-Lingual Voice Conversion With Linguistics Losses to Reduce Foreign Accents},
  year={2023},
  volume={31},
  pages={1916-1926},
  doi={10.1109/TASLP.2023.3271107}
}

@ARTICLE{saito2021nonparallel,
  author={Saito, Yuki and Nakamura, Taiki and Ijima, Yusuke and Nishida, Kyosuke and Takamichi, Shinnosuke},
  journal={Acoustical Science and Technology},
  title={Non-parallel and many-to-many voice conversion using variational autoencoders integrating speech recognition and speaker verification},
  year={2021},
  volume={42},
  number={1},
  pages={1-11},
  doi={10.1250/ast.42.1}
}

@misc{liu2024,
      title={Zero-shot Voice Conversion with Diffusion Transformers}, 
      author={Songting Liu},
      year={2024},
      eprint={2411.09943},
      archivePrefix={arXiv},
      primaryClass={cs.SD},
      url={https://arxiv.org/abs/2411.09943}, 
}

@INPROCEEDINGS{pasad2021layerwise,
  author={Pasad, Ankita and Chou, Ju-Chieh and Livescu, Karen},
  booktitle={2021 IEEE Automatic Speech Recognition and Understanding Workshop (ASRU)}, 
  title={Layer-Wise Analysis of a Self-Supervised Speech Representation Model}, 
  year={2021},
  volume={},
  number={},
  pages={914-921},
  doi={10.1109/ASRU51503.2021.9688093}}

\end{document}